\documentclass[letterpaper,10 pt,conference]{ieeeconf}
\IEEEoverridecommandlockouts
\usepackage{graphicx}

\usepackage{amsmath}
\usepackage{mathrsfs}
\usepackage{amssymb}                           
\usepackage[dvips]{epsfig}
\usepackage{multirow}
\usepackage[noadjust]{cite}

\usepackage{epstopdf}
\usepackage{color}
\usepackage{todonotes}
\usepackage{url}
\usepackage{booktabs}
\usepackage{tabularx}

\everymath{\displaystyle}

\newcommand{\real}{\ensuremath{\mathbb{R}}}
 
\newcommand{\A}{\ensuremath{\mathcal{A}}}

\DeclareMathOperator{\dom}{dom}

\makeatletter
\newcommand{\customlabel}[2]{%
\protected@write \@auxout {}{\string \newlabel {#1}{{#2}{}}}}
\makeatother

\newtheorem{theorem}{\textbf{Theorem}}

\newtheorem{assumption}{\textbf{Assumption}}

\newtheorem{lemma}{Lemma}

\newenvironment{proof}{\noindent\textit{Proof.}}{\hfill $\square$\\}

\title{A Hybrid Controller for Obstacle Avoidance\\
in an $n$-dimensional Euclidean Space}
\author{Soulaimane Berkane, Andrea Bisoffi, Dimos V. Dimarogonas
\thanks{This research was supported in part by the Swedish Research Council (VR), the European Research Council (ERC) through ERC StG BUCOPHSYS, the Swedish Foundation  for Strategic Research (SSF), the EU H2020 Co4Robots project, and the Knut and Alice Wallenberg Foundation (KAW).
The authors are with the Department of Automatic Control, KTH Royal Institute of Technology, Sweden. \texttt{berkane@kth.se} (S. Berkane),  \texttt{bisoffi@kth.se} (A. Bisoffi), \texttt{dimos@kth.se} (D. V. Dimarogonas).}}
\date{September 2018}
\begin{document}

\maketitle
\begin{abstract}
For a vehicle moving in an $n$-dimensional Euclidean space,
we present a construction of a hybrid feedback that guarantees both global asymptotic stabilization of a reference position and avoidance of an obstacle corresponding to a bounded spherical region. The proposed hybrid control algorithm switches between two modes of operation: stabilization (motion-to-goal) and avoidance (boundary-following). The geometric construction of the flow and jump sets of the hybrid controller, exploiting a hysteresis region, guarantees robust switching (chattering-free) between the stabilization and avoidance modes. Simulation results illustrate the performance of the proposed hybrid control approach for a 3-dimensional scenario.
\end{abstract}

\section{Introduction}
The obstacle avoidance problem is a long lasting problem that has attracted the attention of the robotics and control communities for decades. In a typical robot navigation scenario, the robot is required to reach a given goal (destination) while avoiding to collide with a set of obstacle regions in the workspace. Since the pioneering work by Khatib \cite{khatib1986real} and the seminal work by Koditscheck and Rimon \cite{koditschek1990robot}, artificial potential fields or navigation functions have been widely used in the literature, see. {\it e.g.,} \cite{khatib1986real,koditschek1990robot,dimarogonas2006feedback,filippidis2013navigation}, to deal with the obstacle avoidance problem. The idea is to generate an artificial potential field that renders the goal attractive and the obstacles repulsive. Then, by considering trajectories that navigate along the negative gradient of the potential field, one can ensure that the system will reach the desired target from all initial conditions except from a set of measure zero. 
This is a well known topological obstruction to global asymptotic stabilization by continuous time-invariant feedback when the free state space is not diffeomorphic to a Euclidean space, see, e.g., \cite[Thm.~2.2]{wilson1967structure}. This topological obstruction occurs then also in the navigation transform \cite{loizou2017navigation} and (control)-barrier-function approaches \cite{prajna2007framework,wieland2007constructive,romdlony2016stabilization,ames2017control}.

To deal with such a limitation, the authors in \cite{sanfelice2006robust} have proposed a hybrid state feedback controller to achieve robust global asymptotic regulation, in $\mathbb{R}^2$, to a target while avoiding an obstacle. This approach has been exploited in \cite{poveda2018hybrid} to steer a planar vehicle to the source of an unknown but measurable signal while avoiding an obstacle. In \cite{braun2018unsafe}, a hybrid control law has been proposed to globally asymptotically stabilize a class of linear systems while avoiding an unsafe single point in $\mathbb{R}^n$.

In this work, we propose a hybrid control algorithm for the global asymptotic stabilization of a single-integrator system that is guaranteed to avoid a non-point spherical obstacle.
Our approach considers trajectories in an $n-$dimensional Euclidean space and we resort to tools from higher-dimensional geometry \cite{meyer2000matrix} to provide a construction of the flow and jump sets where the different modes of operation of the hybrid controller are activated. 

Our proposed hybrid algorithm employs a hysteresis-based switching between the avoidance controller and the stabilizing controller in order to guarantee forward invariance of the obstacle-free region (related to safety) and global asymptotic stability of the reference position. The parameters of the hybrid controller can be tuned so that the hybrid control law matches the stabilizing controller in arbitrarily large subsets of the obstacle-free region.
 
Preliminaries are in Section~\ref{section:preliminaries}, the problem is formulated in Section~\ref{section:problem}, and our solution is in Sections~\ref{section:controller}-\ref{section:main}, with a numerical exemplification in Section~\ref{section:example}. All the proofs of the intermediate lemmas are in the appendix.

%%%%%%%%%%%%%%%%%%%%%%%%%%%%%%%%%%%
\section{Preliminaries}
\label{section:preliminaries}
Throughout the paper, $\mathbb{R}$ denotes the set of real numbers, $\mathbb{R}^n$ is the $n$-dimensional Euclidean space and $\mathbb{S}^n$ is the $n$-dimensional unit sphere embedded in $\mathbb{R}^{n+1}$. The Euclidean norm of $x\in\mathbb{R}^n$ is defined as $\|x\|:=\sqrt{x^\top x}$ and  the geodesic distance between two points $x$ and $y$ on the sphere $\mathbb{S}^n$ is defined by $\mathbf{d}_{\mathbb{S}^n}(x,y):=\arccos(x^\top y)$ for all $x,y\in\mathbb{S}^n$. The closure, interior and boundary of a set $\mathcal{A}\subset\mathbb{R}^n$ are denoted as  $\overline{\mathcal{A}}, \mathcal{A}^\circ$ and $\partial\mathcal{A}$, respectively. The relative complement of a set $\mathcal{B}\subset\mathbb{R}^n$ with respect to a set $\mathcal{A}$ is denoted by $\mathcal{A}\setminus\mathcal{B}$ and contains the elements of $\mathcal{A}$ which are not in $\mathcal{B}$. % 
%%%%%%%%%%%%%%%%%%%%%%%%%%%%%%%%%%%%%%%%%%%%
Given a  nonzero vector $z\in\mathbb{R}^n\setminus\{0\}$, we define the maps:
\begin{equation}
\label{eq:proj-refl-maps}
\pi^\parallel(z):=\tfrac{zz^\top}{\|z\|^2},\, \pi^\perp(z):=\!I_n\!-\tfrac{zz^\top}{\|z\|^2},\, \rho^\perp(z)=\!I_n\!-2\tfrac{zz^\top}{\|z\|^2}
\end{equation} 
where $I_n$ is the $n\times n$ identity matrix. The map $\pi^\parallel(\cdot)$ is the parallel projection map, $\pi^\perp(\cdot)$ is the orthogonal projection map \cite{meyer2000matrix}, and $\rho^\perp(\cdot)$ is the reflector map (also called Householder transformation). Consequently, for any $x\in\mathbb{R}^n$, the vector $\pi^\parallel(z)x$ corresponds to the projection of $x$ onto the line generated by $z$, $\pi^\perp(z)x$ corresponds to the projection of $x$ onto the hyperplane orthogonal to $z$ and $\rho^\perp(z)x$ corresponds to the reflection of $x$ about the hyperplane orthogonal to $z$. For each $z\in\real^n \setminus\{ 0\}$, some useful properties of these maps follow:
\begin{align}
    \label{eq:propLine1}
    \pi^\parallel(z)z&=z,&\pi^\perp(z)\pi^\perp(z)&=\pi^\perp(z),\\
    \label{eq:propLine2}
    \pi^\perp(z)z&=0,&\pi^\parallel(z)\pi^\parallel(z)&=\pi^\parallel(z), \\
    \label{eq:propLine3}
    \rho^\perp(z)z&=-z,&\rho^\perp(z)\rho^\perp(z)&=I_n,\\
    \label{eq:propLine4}
    \pi^\perp(z)\pi^\parallel(z)&=0,&\pi^\perp(z)+\pi^\parallel(z)&=I_n, \\
    \label{eq:propLine5}
    \pi^\parallel(z)\rho^\perp(z)&=-\pi^\parallel(z),& 2\pi^\perp(z)-\rho^\perp(z)&=I_n,\\
    \label{eq:propLine6}
    \pi^\perp(z)\rho^\perp(z)&=\pi^\perp(z),& 2\pi^\parallel(z)+\rho^\perp(z)&=I_n.
\end{align}
We define for $z\in\real^n\setminus\{ 0\}$
and $\theta\in\real$ the parametric map 
\begin{align}
\label{eq:def:piTheta}
    \pi^\theta(z):=\cos^2(\theta)\pi^\perp(z)-\sin^2(\theta)\pi^\parallel(z).
\end{align}
\begin{figure}
    \centering
    \includegraphics[scale=0.42]{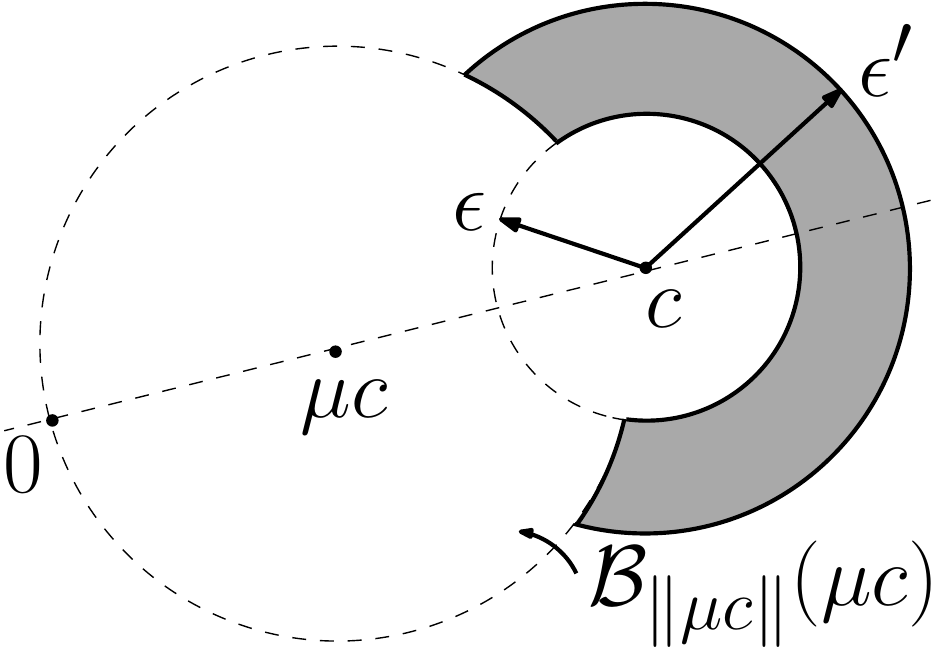}
    \caption{The helmet region (dark grey) defined in \eqref{eq:helmet}.}
    \label{fig:helmet}
\end{figure}%
In~\eqref{eq:def:ball}--\eqref{eq:helmet}, we define for $v\in\real^n\setminus\{ 0\}$ some geometric subsets of $\mathbb{R}^n$, which are described after~\eqref{eq:helmet}:
\begin{align}
    \label{eq:def:ball}
    \mathcal{B}_\epsilon(c)&:=\{x\in\mathbb{R}^n: \|x-c\|\leq\epsilon\}, \\
    \label{eq:def:line}
    \mathcal{L}(c,v)&:=\{x\in\mathbb{R}^n: x=c+\lambda v, \lambda\in\mathbb{R}\}, \\
    \label{eq:def:plane}
    \mathcal{P}^{\bigtriangleup}(c,v)&:=\{x\in\mathbb{R}^n: v^\top(x-c)\bigtriangleup 0\},\\
    \label{eq:def:cone}
    \mathcal{C}^{\bigtriangleup}(c,v,\theta)&:=\{x\in\mathbb{R}^n:\!(x\!-c)^\top\!\pi^\theta\!(v)(x\!-c)\!\bigtriangleup\!0\}\\
    &\!\!\!\!\!\!\!\!\!\!\!\!\!\!\!\!\!\!\!\!=\{x\in\mathbb{R}^n:\cos^2(\theta)\|v\|^2\|x-c\|^2\!\bigtriangleup\! (v^\top(x-c))^2\} \nonumber\\
    \label{eq:def:half_cone}
    \mathcal{C}_{\bigtriangledown}^{\bigtriangleup}(c,v,\theta)&:=\mathcal{C}^{\bigtriangleup}(c,v,\theta)\cap\mathcal{P}^{\bigtriangledown}(c,v),\\
    \label{eq:helmet}
\mathcal{H}(c,\epsilon,\epsilon^\prime,\mu)&:=\overline{\mathcal{B}_{\epsilon^\prime}(c)\setminus\mathcal{B}_{\epsilon}(c)\setminus\mathcal{B}_{\|\mu c\|}(\mu c)},
\end{align}
where the symbols $\bigtriangleup$ and $\bigtriangledown$ can be selected as $\bigtriangleup\in\{=,<,>,\leq,\geq\}$ and $\bigtriangledown\in\{<,>,\leq,\geq\}$. 
The set $\mathcal{B}_\epsilon(c)$ in~\eqref{eq:def:ball} is the \emph{ball} centered at $c\in\mathbb{R}^n$ with radius $\epsilon$. 
The set $\mathcal{L}(c,v)$ in~\eqref{eq:def:line} is the $1-$dimensional \emph{line} passing by the point $c\in\mathbb{R}^n$ and with direction parallel to $v$. 
The set $\mathcal{P}^=(c,v)$ in~\eqref{eq:def:plane} is the $(n-1)-$dimensional \emph{hyperplane} that passes through a point $c\in\mathbb{R}^n$ and has normal vector $v$. 
The hyperplane $\mathcal{P}^=(c,v)$ divides the Euclidean space $\mathbb{R}^n$ into two closed sets $\mathcal{P}^{\geq}(c,v)$ and $\mathcal{P}^\leq(c,v)$. 
The set $\mathcal{C}^=(c,v,\theta)$ in~\eqref{eq:def:cone} is the right circular \emph{cone} with vertex at $c\in\mathbb{R}^n$, axis parallel to $v$ and aperture $2\theta$.
The set $\mathcal{C}^{\bigtriangleup}(c,v,\theta)$ in~\eqref{eq:def:cone} with $\leq$ as $\bigtriangleup$ (or $\geq$ as $\bigtriangleup$, respectively) is the region inside (or outside, respectively) the cone $\mathcal{C}^=(c,v,\theta)$.
The plane $\mathcal{P}^=(c,v)$ divides the conic region $\mathcal{C}^{\bigtriangleup}(c,v,\theta)$ into two regions $\mathcal{C}^{\bigtriangleup}_{\leq}(c,v,\theta)$ and $\mathcal{C}^{\bigtriangleup}_{\geq}(c,v,\theta)$ in~\eqref{eq:def:half_cone}. 
The set $\mathcal{H}(c,\epsilon,\epsilon^\prime,\mu)$ in~\eqref{eq:helmet} is called a {\it helmet} and is obtained by removing from the spherical shell (annulus) $\mathcal{B}_{\epsilon^\prime}(c)\setminus\mathcal{B}_{\epsilon}(c)$ the portion contained in the ball $\mathcal{B}_{\|\mu c\|}(\mu c)$, see Fig.~\ref{fig:helmet}. The following geometric fact will be used. 
%%%%%%%%%%%%%%%%%%%%%%%%%%%%%%%%%%%
\begin{lemma}\label{lemma:cones}
Let $c\in\mathbb{R}^n$ and $v_1,v_2\in\mathbb{S}^{n-1}$ be some arbitrary unit vectors such that $\mathbf{d}_{\mathbb{S}^{n-1}}(v_1,v_2)=\theta$ for some $\theta\in(0,\pi]$. Let $\psi_1,\psi_2\in[0,\pi]$ such that $\psi_1+\psi_2<\theta<\pi-(\psi_1+\psi_2)$. Then
\begin{align*}
    \mathcal{C}^{\leq}(c,v_1,\psi_1)\cap\mathcal{C}^{\leq}(c,v_2,\psi_2)=\{c\}.
\end{align*}
\end{lemma}
Finally, we consider in this paper hybrid dynamical systems \cite{goebel2012hybrid}, described through constrained differential and difference inclusions for state $X \in \real^n$:
\begin{equation}
\label{Hybrid:general}
\begin{cases}
\dot X\in\mathbf{F}(X), &X\in\mathcal{F},\\
X^+\in \mathbf{J}(X),  & X\in\mathcal{J}.
\end{cases}
\end{equation}
The data of the hybrid system \eqref{Hybrid:general} (i.e., the \textit{flow set} $\mathcal{F}\subset\mathbb{R}^n$, the \textit{flow map} $\mathbf{F}:\mathbb{R}^n\rightrightarrows\mathbb{R}^n$, the \textit{jump set} $\mathcal{J}\subset\mathbb{R}^n$, the \textit{jump map} $\mathbf{J}:\mathbb{R}^n\rightrightarrows\mathbb{R}^n$) is denoted as $\mathscr{H}=(\mathbf{\mathcal{F}},\mathbf{F},\mathcal{J},\mathbf{J})$.
%%%%%%%%%%%5
\section{Problem Formulation}\label{section:problem}
We consider a vehicle moving in the $n$-dimensional Euclidean space according to the following single integrator dynamics:
\begin{align}
    \dot x=u
\end{align}
where $x\in\mathbb{R}^n$ is the state of the vehicle and $u\in\mathbb{R}^n$ is the control input. We assume that in the workspace there exists an obstacle considered as a spherical region $\mathcal{B}_\epsilon(c)$ centered at $c\in\mathbb{R}^n$ and with radius $\epsilon>0$. The vehicle needs to avoid the obstacle while stabilizing its position to a given reference. Without loss of generality we consider $n\geq 2$ and take  our reference %(desired) 
position at $x=0$ (the origin)\footnote{ \label{footnote:n=1} For $n=1$ (i.e., the state space is a line), global asymptotic stabilization with obstacle avoidance is physically impossible to solve via any feedback.}. 
\begin{assumption}\label{assumption:obstacle}
$\|c\|>\epsilon>0$.
\end{assumption}
Assumption \ref{assumption:obstacle} requires that the reference position $x=0$ is not inside the obstacle region, otherwise the following control objective would not be feasible. Our objective is indeed to design a control strategy for the input $u$ such that:
\begin{itemize}
    \item[i)] the obstacle-free region $\mathbb{R}^n\setminus\mathcal{B}_{\epsilon}(c)$ is forward invariant;
    \item[ii)] the origin $x=0$ is globally asymptotically stable;
    \item[iii)] for each $\epsilon^\prime>\epsilon$, there exist controller parameters such that the control law matches, in  $\mathbb{R}^n\setminus\mathcal{B}_{\epsilon^\prime}(c)$, the law $u=-k_0x$  ($k_0>0$) used in the absence of the obstacle. 
    \end{itemize}
Objective i) guarantees that all trajectories of the closed-loop system are safely avoiding the obstacle by remaining outside the obstacle region. Objectives i) and ii), together, can not be achieved using a continuous feedback due to the topological obstruction discussed in the introduction. Objective iii) is the so-called {\it semiglobal preservation} property \cite{braun2018unsafe}. This property is desirable when the original controller parameters are optimally tuned and the controller modifications imposed by the presence of the obstacle should be as minimal as possible. Such a property is also accounted for in the quadratic programming formulation of~\cite[III.A.]{wang2017safety}.
The obstacle avoidance problem described above is solved via a hybrid feedback strategy in Sections~\ref{section:controller}-\ref{section:main}.
\section{Proposed Hybrid Control Algorithm for Obstacle Avoidance}\label{section:controller}
\begin{figure}
    \centering
    \includegraphics[width=\columnwidth]{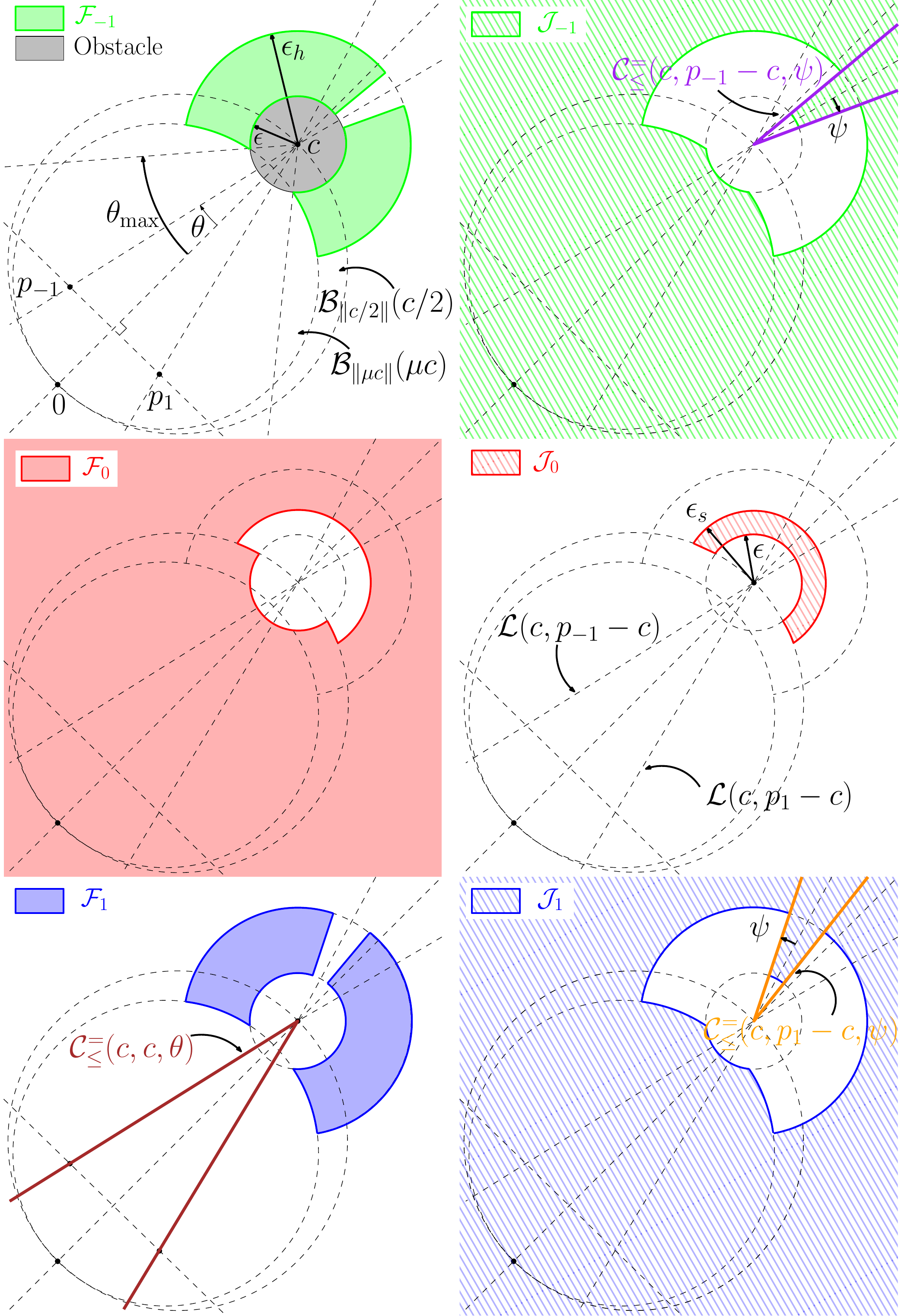}
    \caption{2D illustration of flow and jump sets considered in Sections~\ref{section:controller}-\ref{section:main}.}
    \label{fig:flowAndJumpSets}
\end{figure}

In this section, we propose a hybrid controller that switches suitably between an {\it avoidance} controller and a {\it stabilizing} controller. Let $m\in\{-1,0,1\}$ be a discrete variable dictating the control mode where $m=0$ corresponds to the activation of the stabilizing controller and $|m|=1$ corresponds to the activation of the avoidance controller, which has two configurations $m\in\{-1,1\}$. The proposed control input, depending on both the state $x\in\mathbb{R}^n$ and the control mode $m\in\{-1,0,1\}$, is given by the feedback law
\begin{equation*}\label{eq:u}
\begin{aligned}
    u& =\kappa(x,m):=\begin{cases}
    -k_0 x, & m=0\\
    - k_m \pi^\perp(x-c)(x-p_m),&m \in\{-1,\, 1\}
    \end{cases}
\end{aligned}
\end{equation*}
where $k_m>0$ (with $m\in\{-1,0,1\}$) and $p_m\in\mathbb{R}^n$ (with $m\in\{-1,1\}$) are design parameters. During the stabilization mode ($m=0$), the control input above steers $x$ towards $x=0$. During the avoidance mode ($|m|=1$), the control input above minimizes the distance to the \emph{auxiliary} attractive point $p_m$ {\it while} maintaining a constant distance to the center of the ball $\mathcal{B}_{\epsilon}(c)$, thereby avoiding to hit the obstacle. This is done by projecting the feedback $-k_m(x-p_m)$ on the hyperplane orthogonal to $(x-c)$. This control strategy resembles the well-known path planning Bug algorithms (see, {\it e.g.,} \cite{lumelsky1990incorporating}) where the motion planner switches between motion-to-goal objective and boundary-following objective.
We refer the reader to Fig.~\ref{fig:flowAndJumpSets} from now onward for all of this section.
For $\theta>0$ (further bounded in~\eqref{ineq:parameters}), the points $p_1, p_{-1}$ are selected to lie on the cone\footnote{Following the remark in Footnote~\ref{footnote:n=1}, note that the set $\mathcal{C}^=_\leq(c,c,\theta)\setminus\{c\}$ is nonempty for all $n\geq 2$.} $\mathcal{C}^=_\leq(c,c,\theta)\setminus\{c\}$: 
\begin{align}\label{eq:p-1}
    p_1\in\mathcal{C}^=_\leq(c,c,\theta)\setminus\{c\} \text{ and } p_{-1}:=-\rho^\perp(c)p_1.
\end{align}
Note that, by~\eqref{eq:p-1}, $p_{-1}$ opposes $p_1$ diametrically with respect to  the axis of the cone $\mathcal{C}^=_\leq(c,c,\theta)$ and also belongs to $\mathcal{C}^=_\leq(c,c,\theta)\setminus\{c\}$ as shown in the following lemma.
\begin{lemma}\label{lemma:p-1}
$p_{-1}\in\mathcal{C}^=_\leq(c,c,\theta)\setminus\{c\}.$
\end{lemma}
The logic variable $m$ is selected according to a hybrid mechanism that exploits a suitable construction of the flow and jump sets. This hybrid selection is obtained through the  hybrid dynamical system
\begin{subequations}
\label{eq:hs_1obs}
\begin{align}
&\left\{\begin{aligned}
\dot x&=\kappa(x,m)\\
\dot m&=0
\end{aligned}\right.&&(x,m)\in\bigcup_{m \in \{-1,0,1\}} \!\!\!\! \mathcal{F}_m \times \{m\}\label{eq:hs_1obs:flowMap}\\
&\left\{\begin{aligned}
x^+ &=x\\
m^+ &\in\mathbf{M}(x,m)
\end{aligned}\right.&&(x,m)\in\bigcup_{m \in \{-1,0,1\}} \!\!\!\! \mathcal{J}_m \times \{m\}. \label{eq:hs_1obs:JumpMap}
\end{align}
The flow and jump sets for each mode $m\in\{-1,0,1\}$ are defined as (see~\eqref{eq:helmet} for the definition of the helmet $\mathcal{H}$):
\begin{align}
\label{eq:F0}
& \mathcal{F}_0:=\overline{\mathbb{R}^n\setminus(\mathcal{J}_0\cup\mathcal{B}_{\epsilon}(c))},  & & \\ 
\label{eq:J0}
&\mathcal{J}_0:=\mathcal{H}(c,\epsilon,\epsilon_s,1/2), & &\\
\label{eq:Fm}
&\mathcal{F}_m:=\mathcal{H}(c,\epsilon,\epsilon_h,\mu)\cap\mathcal{C}_\leq^\geq(c,p_m-c,\psi), & & |m|=1,\\
\label{eq:Jm}
&\mathcal{J}_m:=\overline{\mathbb{R}^n\setminus(\mathcal{F}_m\cup\mathcal{B}_{\epsilon}(c))},& & |m|=1,
\end{align}
see their depiction in Fig.~\ref{fig:flowAndJumpSets}, and the (set-valued) jump map is defined  as
\begin{align}
    \mathbf{M}(x,0)&\!:=\left\{m^\prime\!\in\!\{-1,1\}\colon x\in\mathcal{C}^{\geq}(c,p_{m^\prime}\!-c,\bar\psi)\right\} \label{eq:hs_1obs:M(x,0)}\\
    \mathbf{M}(x,m)&\!:= 0, \quad \text{ for } m\in \{-1,1\},
\end{align}
\end{subequations}
%%%%%%%%%%%%%%%%%%%%%%%%%%%%%
%%%%%%%%%%%%%%%%%%%%
where $\epsilon_s$, $\epsilon_h$, $\theta$, $\psi$, $\bar \psi$ are design parameters selected later as in Assumption~\ref{assumption:parameters}. Before we state our main result, a discussion motivating the above construction of flow and jump sets is in order. 

During the stabilization mode $m=0$, the closed-loop system should not flow when $x$ is close enough to the surface of the obstacle region $\mathcal{B}_{\epsilon}(c)$ and the vector field $-k_0x$ points inside $\mathcal{B}_{\epsilon}(c)$. Indeed, by computing the time derivative of $\|x-c\|^2$, we can obtain the set where the stabilizing vector field $-k_0x$ causes a decrease in the distance $\|x-c\|^2$ to the centre of the obstacle region $\mathcal{B}_\epsilon(c)$. This set is characterized by the inequality
\begin{align}
\label{eq:dist_to_c_decreases}
-k_0 x^\top(x-c)\leq 0 \Longleftrightarrow 
\left\|x-{c}/{2}\right\|^2\geq\left\|{c}/{2}\right\|^2.
\end{align}
The closed set in~\eqref{eq:dist_to_c_decreases} corresponds to the region outside the ball $\mathcal{B}_{\|c/2\|}(c/2)$. Therefore, to keep the vehicle safe during the stabilization mode, we define around the obstacle a helmet region $\mathcal{H}(c,\epsilon,\epsilon_s,1/2)$, which is used as the jump set $\mathcal{J}_0$ in \eqref{eq:J0}. In other words, if during the stabilization mode the vehicle hits this {\it safety helmet}, then the controller jumps to avoidance mode. The amount $\epsilon_s-\epsilon$ represents the thickness of the safety helmet that defines the jump set $\mathcal{J}_0$.

During the avoidance mode $|m|=1$, we want our controller to slide on the helmet $\mathcal{H}(c,\epsilon,\epsilon_h,\mu)$ while maintaining a constant distance to the center $c$. Note that, with $\epsilon_h>\epsilon_s$ and $\mu<1/2$, the helmet $\mathcal{H}(c,\epsilon,\epsilon_h,\mu)$ (see also Fig.~\ref{fig:helmet}) is an {\it inflated} version of the helmet $\mathcal{H}(c,\epsilon,\epsilon_s,1/2)$ and creates a hysteresis region useful to prevent infinitely many consecutive jumps (Zeno behavior). Let us then characterize in the following lemma the equilibria of the avoidance vector field $\kappa(x,m)= - k_m \pi^\perp(x-c)(x-p_m)$ ($|m|=1$).
\begin{lemma}\label{lemma:equilibria}
For each $x\in\real^n \setminus \{c\}$ and $m \in\{-1, 1\}$, $\pi^\perp(x-c)(x-p_m)=0$ if and only if 
$x\in\mathcal{L}(c,p_m-c)$.
\end{lemma}
Since we want the trajectories to leave the set $\mathcal{F}_m$ during the avoidance mode, it is necessary to select the point $p_m$ and the flow set $\mathcal{F}_m$ such that $\mathcal{L}(c,p_m-c)\cap\mathcal{F}_m = \emptyset$ for  each $m\in\{-1,1\}$, otherwise trajectories can stay in the avoidance mode indefinitely. This motivates the intersection with the conic region in~\eqref{eq:Fm} and Lemma~\ref{lemma:empty1}, in view of which we pose the following assumption.
\begin{assumption}\label{assumption:parameters}
The parameters in~\eqref{eq:hs_1obs} are selected as:
\begin{align}
& \epsilon_h \in \big(\epsilon, \sqrt{\epsilon\|c\|}\big) &&\epsilon_s\in(\epsilon, \epsilon_h)
&&\mu \in (\mu_{\min},1/2) \label{ineq:eps_h,eps_s}\\
&\theta\in(0,\theta_{\max})
&& \psi\in(0,\psi_{\max}) \label{ineq:parameters}
&& \bar\psi \in(\psi,\psi_{\max})
\end{align}
where $\mu_{\min}$, $\theta_{\max}$ and $\psi_{\max}$ are defined  as
\begin{align}
&\mu_{\min}:=\frac{1}{2} \frac{\epsilon_h^2+ \| c \|^2 - 2 \epsilon \| c \|}{\| c \|^2 - \epsilon \| c \|} \in (0,{1}/{2}),\\
&\theta_{\max}:=\arccos\left(\frac{\epsilon_h^2+\| c \|^2(1-2 \mu)}{2\epsilon  \| c\|(1-\mu)}\right) \in (0,{\pi}/{2}) \label{eq:theta_max},\\
& \psi_{\max} := \min(\theta,\pi/2-\theta) \in (0,\pi/4).
\label{eq:psi_max}
\end{align}
\end{assumption}
The intervals in~\eqref{ineq:eps_h,eps_s}--\eqref{eq:psi_max} are well defined. They can be checked in this order. The intervals of $\epsilon_h$ and $\epsilon_s$ are well defined by Assumption~\ref{assumption:obstacle}. Then, those of $\mu_{\min}$, $\mu$, $\theta_{\max}$ ($\theta_{\max}>0$ directly from $\mu > \mu_{\min}$), $\theta$, $\psi_{\max}$ and, finally, those of $\psi$ and $\bar \psi$ (corresponding to $0< \psi < \bar \psi < \psi_{\max}$) are also well defined.
\begin{lemma}\label{lemma:empty1}
Under Assumption~\ref{assumption:parameters}, $\mathcal{F}_m\cap\mathcal{L}(c,p_m-c)=\emptyset$, for  $m\in\{-1,1\}$.
\end{lemma}

%%%%%%%%%%%%%%%%%%%%%%%%%%%%%%%%%%%%%%%%%%%%%%%%%%%%%%%%%%%%%%%%%%%%%%%%%
\section{Main Result}\label{section:main}
In this section, we state and prove our main result, which corresponds to the objectives discussed in Section \ref{section:problem}. Let us first write more compactly flow/jump sets and maps:
\begin{align}
\label{eq:hs_1obs:flowJumpSets}
\mathcal{F}:=\!\!\!\! \bigcup_{m \in \{-1,0,1\}} \!\!\!\! & \mathcal{F}_m \times \{m\},\, \mathcal{J}:=\!\!\!\! \bigcup_{m \in \{-1,0,1\}} \!\!\!\!\mathcal{J}_m \times \{m\}\\
    (x,m) & \mapsto \mathbf{F}(x,m) := (\kappa(x,m),0),\\
    (x,m) & \mapsto \mathbf{J}(x,m) := (x,\mathbf{M}(x,m)).
\end{align}
The mild regularity conditions satisfied by the hybrid system~\eqref{eq:hs_1obs}, as in the next lemma, guarantee the applicability of many results in the proof of our main result.
\begin{lemma}
\label{lemma:hbc}
The hybrid system with data $(\mathcal{F},\mathbf{F},\mathcal{J},\mathbf{J})$ satisfies the hybrid basic conditions in~\cite[Ass.~6.5]{goebel2012hybrid}.
\end{lemma}
Let us define the obstacle-free set $\mathcal{K}$ and the attractor $\mathcal{A}$ as:
\begin{equation}
\label{eq:KandA}
\mathcal{K}:=\overline{\mathbb{R}^n\setminus\mathcal{B}_\epsilon(c)}\times\{-1,0,1\},\quad \A:=\{0\}\times\{0\}.
\end{equation}
Our main result is given in the following theorem.
\begin{theorem}\label{theorem:invariance}
Consider the hybrid system \eqref{eq:hs_1obs} under Assumptions~\ref{assumption:obstacle}-\ref{assumption:parameters}. Then, 
\begin{itemize}
    \item[i)] all maximal solutions do not have finite escape times, are complete in the ordinary time direction, and the obstacle-free set $\mathcal{K}$ in~\eqref{eq:KandA} is forward invariant;
    \item[ii)] the set $\A$ in~\eqref{eq:KandA} is %uniformly
    globally asymptotically stable;
    \item[iii)] for each $\epsilon^\prime>\epsilon$, it is possible to tune the hybrid controller parameters so that the resulting hybrid feedback law matches, in $\mathbb{R}^n\setminus\mathcal{B}_{\epsilon^\prime}(c)$, the law $u=-k_0x$.
\end{itemize}
\end{theorem}
Theorem \ref{theorem:invariance} shows that the three objectives discussed in Section \ref{section:problem} are fulfilled. 
\subsection{Proof of Theorem \ref{theorem:invariance}}
\begin{table}
\[
\begin{array}{ll}
\toprule
\text{Set to which $x$ belongs}& \mathbf{T}_{\mathcal{F}_0}(x)\\
\midrule
\partial\mathcal{B}_\epsilon(c)\cap\mathcal{B}^\circ_{\|c/2\|}(c/2) & \mathcal{P}^\geq(0,x-c)\\
\partial\mathcal{B}_{\epsilon_s}(c)\setminus\mathcal{B}_{\|c/2\|}(c/2)& \mathcal{P}^\geq(0,x-c)\\
(\partial\mathcal{B}_{\|c/2\|}(c/2)\cap\mathcal{B}^\circ_{\epsilon_s}(c))\setminus\mathcal{B}_\epsilon(c) & \mathcal{P}^\leq(0,x-c/2)\\
\partial\mathcal{B}_{\epsilon}(c)\cap\partial\mathcal{B}_{\|c/2\|}(c/2) & \mathcal{P}^\geq(0,x-c)\cap\mathcal{P}^\leq(0,x-c/2)\\
\partial\mathcal{B}_{\|c/2\|}(c/2)\cap\partial\mathcal{B}_{\epsilon_s}(c) & \mathcal{P}^\geq(0,x-c)\cup\mathcal{P}^\leq(0,x-c/2)\\
\bottomrule
\end{array}
\]
\[
\begin{array}{ll}
\toprule
\text{Set to which $x$ belongs}& \mathbf{T}_{\mathcal{F}_{\bar m}}(x) \\
\midrule
\partial\mathcal{B}_\epsilon(c)\!\setminus\!\mathcal{B}_{\|\mu c\|}(\mu c)\!\setminus\!\mathcal{C}^\leq(c,p_{\bar m}\!-\!c,\psi) & \mathcal{P}^\geq(0,x-c)\\
\partial\mathcal{B}_{\epsilon_h}\!(c)\!\setminus\!\mathcal{B}_{\|\mu c\|}\!(\mu c)\!\setminus\!\mathcal{C}^\leq(c,p_{\bar m}\!-\!c,\psi)
& \mathcal{P}^\leq(0,x-c)\\
\partial\mathcal{B}_{\|\mu c\|}(\mu c)\cap\mathcal{B}^\circ_{\epsilon_h}(c)\setminus\mathcal{B}_{\epsilon}(c) & \mathcal{P}^\geq(0,x-\mu c)\\
\mathcal{C}^=_{\le}(c,p_{\bar m}-c,\psi)\cap\mathcal{B}^\circ_{\epsilon_h}(c)\setminus\mathcal{B}_{\epsilon}(c) & \mathcal{P}^\geq(0,n_{\bar m}(x))\\
\partial\mathcal{B}_{\epsilon}(c)\cap\partial\mathcal{B}_{\|\mu c\|}(\mu c) & \mathcal{P}^\geq(0,x\!-\!c)\!\cap\!\mathcal{P}^\geq(0,x-\mu c)\\
\partial\mathcal{B}_{\epsilon_h}(c)\cap\partial\mathcal{B}_{\|\mu c\|}(\mu c) & \mathcal{P}^\leq(0,x\!-\!c)\!\cap\!\mathcal{P}^\geq(0,x-\mu c)\\
\partial\mathcal{B}_{\epsilon}(c)\cap\mathcal{C}^=(c,p_{\bar m}-c,\psi) & \mathcal{P}^\geq(0,x\!-\!c)\!\cap\!\mathcal{P}^\geq(0,n_{\bar m}(x))\\
\partial\mathcal{B}_{\epsilon_h}(c)\cap\mathcal{C}^=(c,p_{\bar  m}-c,\psi) & \mathcal{P}^\leq(0,x\!-\!c)\!\cap\!\mathcal{P}^\geq(0,n_{\bar m}(x))\\
\bottomrule
\end{array}
\]
\caption{Points $(x,m)$ and their tangent cones ($\bar m$ is either $-1$ or $1$ and $n_{\bar m}(x):=\pi^{\psi}(p_{\bar m}-c)(x-c)$).}
\label{eq:tangent_cone}
\end{table}
%%%%%%%%%%%%%%%%%%%%%%%%%%%%%%%%%%%%%%%%
To prove item i), we resort to~\cite[Thm.~4.3]{chai2018forward}. We first establish for $\mathscr{H}$ in~\eqref{eq:hs_1obs} the relationships invoked in~\cite[Thm.~4.3]{chai2018forward}, and we refer the reader to Fig.~\ref{fig:flowAndJumpSets} for a two-dimensional visualization. In particular, the boundary of the flow set $\mathcal{F}$ is given by $\partial\mathcal{F}=\{(x,m):x\in\partial\mathcal{F}_m\}$, where the sets $\partial\mathcal{F}_0$ and $\partial\mathcal{F}_m, m\in\{-1,1\}$, are 
\begin{align*}
\nonumber
     \partial\mathcal{F}_0&=\big(\partial\mathcal{B}_\epsilon(c)\cap\mathcal{B}_{\|c/2\|}(c/2)\big)
     \cup\big(\partial\mathcal{B}_{\epsilon_s}(c)\setminus\mathcal{B}_{\|c/2\|}(c/2)\big)\\
     &\quad\cup\big((\partial\mathcal{B}_{\|c/2\|}(c/2)\cap\mathcal{B}_{\epsilon_s}(c))\setminus\mathcal{B}_\epsilon(c)\big),\\
\nonumber
    \partial\mathcal{F}_m&=\big((\partial\mathcal{B}_\epsilon(c)\cup\partial\mathcal{B}_{\epsilon_h}(c))\setminus\mathcal{B}_{\|\mu c\|}(\mu c)\setminus\mathcal{C}^\leq_\le(c,p_m-c,\psi)\big)\\
    &\cup\big((\partial\mathcal{B}_{\|\mu c\|}(\mu c)\cup\mathcal{C}^=_\leq(c,p_m-c,\psi))\cap\mathcal{B}_{\epsilon_h}(c)\setminus\mathcal{B}^\circ_{\epsilon}(c)\big).
\end{align*}
The tangent cone\footnote{For the definition of tangent cone, see~\cite[Def.~5.12 and Fig.~5.4]{goebel2012hybrid}.}, evaluated at the boundary of $\mathcal{F}$, is given in Table~\ref{eq:tangent_cone}. Consider $m=0$ and let $z:=\kappa(x,0)=-k_0x$. 
If $x\in\partial\mathcal{B}_\epsilon(c)\cap\mathcal{B}^\circ_{\|c/2\|}(c/2)$ then one has $(x-c)^\top z=-k_0x^\top(x-c)>0$ (since $x \in \mathcal{B}^\circ_{\| c/2 \|}(c/2)$,  see~\eqref{eq:dist_to_c_decreases}), i.e., $z\in\mathcal{P}^>(0,x-c)$. If $x\in(\partial\mathcal{B}_{\|c/2\|}(c/2)\cap\mathcal{B}^\circ_{\epsilon_s}(c))\setminus\mathcal{B}_{\epsilon}(c)$ then one has $(x-c/2)^\top z=-k_0x^\top(x-c/2)=-k_0x^\top c/2=-k_0\|x\|^2/2 \le 0$ since $x^\top(x-c)=0$ from $\|x-c/2\|=\|c/2\|$. Then, $z\in\mathcal{P}^\le (0,x-c/2)$. 
If $x\in\partial\mathcal{B}_{\epsilon}(c)\cap\partial\mathcal{B}_{\|c/2\|}(c/2)$ or $x\in\partial\mathcal{B}_{\|c/2\|}(c/2)\cap\partial\mathcal{B}_{\epsilon_s}(c)$ then $z^\top(x-c)=0$ and $z^\top(x-c/2)=-k_0\|x\|^2/2\leq 0$ showing, respectively, that $z\in\mathcal{P}^\geq(0,x-c)\cap\mathcal{P}^\leq(0,x-c/2)$. 
Finally, if $x\in\partial\mathcal{B}_{\epsilon_s}(c)\setminus\mathcal{B}_{\|c/2\|}(c/2)$, then one has $(x-c)^\top z=-k_0x^\top(x-c)< 0$ (since $x\notin \mathcal{B}_{\| c / 2\|}(c/2)$), i.e., $z\in\mathcal{P}^<(0,x-c)$. Let $\mathcal{L}_0:=\partial\mathcal{B}_{\epsilon_s}(c)\setminus\mathcal{B}_{\|c/2\|}(c/2)$. Therefore, by all the previous arguments,
\begin{equation}
\label{eq:tgConeBndF0}
    \begin{aligned}
    x\in\mathcal{L}_0 & \implies \kappa(x,0)\cap\mathbf{T}_{\mathcal{F}_0}(x)=\emptyset\\
    x\in\partial\mathcal{F}_0\setminus\mathcal{L}_0 & \implies \kappa(x,0)\subset\mathbf{T}_{\mathcal{F}_0}(x).
    \end{aligned}
\end{equation}
Consider then $m\in\{-1,1\}$ and let now $z:=\kappa(x,m)=-k_m\pi^\perp(x-c)(x-p_m)$. 
If $x\in\partial\mathcal{B}_{\epsilon}(c)$ or $x\in\partial\mathcal{B}_{\epsilon_h}(c)$ then one has $(x-c)^\top z=-k_m(x-c)^\top\pi^\perp(x-c)(x-p_m)=0$, which implies that both $z\in\mathcal{P}^\geq(0,x-c)$ and $z\in\mathcal{P}^\leq(0,x-c)$. 
Define $n_m(x):=\pi^{\psi}(p_m-c)(x-c)$, which is a normal vector to the cone $\mathcal{C}^=(c,p_m-c,\psi)$ at $x$.
If $x\in\mathcal{C}^=_\leq(c,p_m-c,\psi)$, then\footnote{Each (in)equality is obtained thanks to the relationship reported over it. \label{note:overset}}
\begin{align*}
    &n_m(x)^\top z=-k_m n_m(x)^\top\pi^\perp(x-c)(x-p_m)\\
    &\overset{\eqref{eq:propLine2}}{=}k_m(x-c)^\top\pi^{\psi}(p_m-c) \pi^\perp(x-c)(p_m-c)\\
    &\overset{\eqref{eq:def:piTheta},\eqref{eq:propLine4}}{=}\!k_m(x-c)^\top\!(\pi^\perp\!(p_m-c)\! -\!\sin^2(\psi)I_n )\pi^\perp\!(x-c)(p_m-c)\\
    &\overset{\eqref{eq:propLine2}}{=}k_m(x-c)^\top\pi^\perp(p_m-c)\pi^\perp(x-c)(p_m-c)\\
    &\overset{\eqref{eq:propLine4}}{=} k_m(x-c)^\top\pi^\perp(p_m-c)\big(I_n - \pi^\parallel(x-c)\big)(p_m-c)\\
    &\overset{\eqref{eq:propLine2}}{=}-k_m(x-c)^\top\pi^\perp(p_m-c)\pi^\parallel(x-c)(p_m-c)\\
    &\overset{\eqref{eq:proj-refl-maps}}{=}-k_m\frac{(x-c)^\top\pi^\perp(p_m-c)(x-c)}{\|x-c\|^2}(x-c)^\top(p_m-c)\geq 0
\end{align*}
where the last bound follows from $\pi^\perp(p_m-c)$ positive semidefinite and $(x-c)^\top(p_m-c)\leq 0$ (since $x\in\mathcal{C}^=_\leq(c,p_m-c,\psi)\subset\mathcal{P}^{\leq}(c,p_m-c)$). Hence, $z\in\mathcal{P}^\geq(0,n_m(x))$. Finally, let $x\in\partial\mathcal{B}_{\|\mu c\|}(\mu c)\cap\mathcal{B}_{\epsilon_h}(c)\setminus\mathcal{B}^\circ_{\epsilon}(c)$. With $\theta_{\max}$ in~\eqref{eq:theta_max}, we have
\begin{subequations}
\label{eq:proofRel}
\begin{align}
    &  0 \le c^\top (c- x) \le \cos(\theta_{\max}) \| c \| \| x- c\| \label{eq:proofRel1} \\
    & |(x-c)^\top (p_m -c)| \le \| x-c \| \| p_m - c\| \label{eq:proofRel2}\\
    & c^\top (p_m - c) = - \cos(\theta) \| c \| \| p_m - c \| \label{eq:proofRel3}
\end{align}
\end{subequations}
where the bounds in~\eqref{eq:proofRel1} follow from \eqref{eq:whenInH(epsilon_h,mu)} in the proof of the previous Lemma~\ref{lemma:empty1}, $\mu<1/2$, and $x\in\partial\mathcal{B}_{\|\mu c\|}(\mu c)\cap\mathcal{B}_{\epsilon_h}(c)\setminus\mathcal{B}^\circ_{\epsilon}(c) \subset \mathcal{H} (c,\epsilon,\epsilon_h, \mu)$; \eqref{eq:proofRel3} follows from $p_m \in \mathcal{C}^=_\leq(c,c,\theta)$ (by~\eqref{eq:p-1} and Lemma~\ref{lemma:p-1}). So %\footnotemark[\ref{note:overset}],
\begin{equation*}
\begin{aligned}
   &(x-\mu c)^\top z=-k_m(x-\mu c)^\top\pi^\perp(x-c)(x-p_m)\\
   &\overset{\eqref{eq:propLine2}}{=}k_m(c-\mu c)^\top\pi^\perp(x-c)(p_m-c)\\
   &\overset{\eqref{eq:proj-refl-maps}}{=}k_m(1-\mu)(c^\top(p_m-c)+\\
   &\qquad c^\top(c-x)(x-c)^\top(p_m-c)/\|x-c\|^2)\\
   &\overset{\eqref{eq:proofRel}}{\leq} k_m(1-\mu)(-\cos(\theta)+\cos(\theta_{\max}))\|c\|\|p_m-c\|<0
\end{aligned}
\end{equation*}
since $k_m>0$, $1-\mu >0$ (from~\eqref{ineq:eps_h,eps_s}) and $\theta < \theta_{\max}$ (from~\eqref{ineq:parameters}). $(x-\mu c)^\top z < 0$ implies then $z\in\mathcal{P}^<(0,x-\mu c)$. Let $\mathcal{L}_m:=\partial\mathcal{B}_{\|\mu c\|}(\mu c)\cap\mathcal{B}_{\epsilon_h}(c)\setminus\mathcal{B}^\circ_{\epsilon}(c)$. Therefore, by all the previous arguments,
\begin{equation}
\label{eq:tgConeBndFm}
    \begin{aligned}
    x\in\mathcal{L}_m & \implies \kappa(x,m)\cap\mathbf{T}_{\mathcal{F}_m}(x)=\emptyset\\
    x\in\partial\mathcal{F}_m\setminus\mathcal{L}_m & \implies \kappa(x,m)\subset\mathbf{T}_{\mathcal{F}_m}(x).
    \end{aligned}
\end{equation}
We can now apply \cite[Thm.~4.3]{chai2018forward}. With $\mathcal{K}$ in~\eqref{eq:KandA}, let $\hat{\mathcal{F}}:=\partial(\mathcal{K}\cap\mathcal{F})\setminus\mathcal{L}$ with $\mathcal{L}:=\{(x,m)\in\partial\mathcal{F}: \mathbf{F}(x,m)\cap\mathbf{T}_{\mathcal{F}}(x,m)=\emptyset\}$. By~\eqref{eq:tgConeBndF0} and \eqref{eq:tgConeBndFm} and $\mathcal{K}\cap\mathcal{F}=\mathcal{F}$, we have $\hat{\mathcal{F}}=\cup_{m=-1,0,1}(\partial\mathcal{F}_m\setminus\mathcal{L}_m)\times\{m\}$
 and $\mathcal{L}=\cup_{m=-1,0,1}\mathcal{L}_m\times\{m\}$. It follows from~\eqref{eq:tgConeBndF0} and \eqref{eq:tgConeBndFm} that for every $(x,m)\in\hat{\mathcal{F}}$, $\mathbf{F}(x,m)\subset\mathbf{T}_{\mathcal{F}}(x,m)$. Also, $\mathbf{J}(\mathcal{K}\cap\mathcal{J})\subset\mathcal{K}$, $\mathcal{F}$ is closed, the map $\mathbf{F}$ satisfies the hybrid basic conditions as proven in Lemma~\ref{lemma:hbc} and it is, moreover, locally Lipschitz since it is continuously differentiable. We conclude then that the set $\mathcal{K}$ is forward pre-invariant \cite[Def.~3.3]{chai2018forward}. In addition, since $\mathcal{L}_0\subset\mathcal{J}_0$ and $\mathcal{L}_m\subset\mathcal{J}_m$ with $m\in\{-1,1\}$, one has $\mathcal{L}\subset\mathcal{J}$. Besides, finite escape times can only occur through flow, and since the sets $\mathcal{F}_{-1}$ and $\mathcal{F}_1$ are bounded by their definitions in~\eqref{eq:Fm}, finite escape times cannot occur for $x \in \mathcal{F}_{-1} \cup \mathcal{F}_1$. They can neither occur for $x \in \mathcal{F}_{0}$ because they would make $x^\top x$ grow unbounded, and this would contradict that $\tfrac{d}{dt}(x^\top x) \le 0$ by the definition of $\kappa(x,0)$ and by~\eqref{eq:hs_1obs:flowMap}. Therefore, all maximal solutions do not have finite escape times. By~\cite[Thm.~4.3]{chai2018forward} again, the set $\mathcal{K}$ is actually forward invariant \cite[Def.~3.3]{chai2018forward}, and solutions are complete.  Finally, we anticipate here a straightforward corollary of completeness and Lemma~\ref{lemma:finiteJumps} below: since the number of jumps is finite by Lemma~\ref{lemma:finiteJumps}, all maximal solutions to~\eqref{eq:hs_1obs} are actually complete in the ordinary time direction.

Now, we will prove item ii) in two steps. First, we prove in the following Lemma~\ref{lemma:GAS_jumpless} that the set $\A$ is globally asymptotically stable for the system without jumps. To this end, the \emph{jumpless system} has data
$
    \mathscr{H}^0 =(\mathbf{F}, \mathcal F, \emptyset, \emptyset )
$
with flow map $\mathbf{F}$ and flow set $\mathcal F$ defined in~\eqref{eq:hs_1obs}. We emphasize that $\mathscr{H}^0$ is obtained in accordance to~\cite[Eqq.~(38)-(39)]{goebel2009hybrid} by identifying \emph{all} jumps with events.
%%%%%%%%%%%%%%%%%%%%%%%%
\begin{lemma}
\label{lemma:GAS_jumpless}
$\A$ in~\eqref{eq:KandA} is globally asymptotically stable for the jumpless hybrid system $\mathscr{H}^0$.
\end{lemma}
Second, we prove in the following Lemma~\ref{lemma:finiteJumps} that the number of jumps is finite for the given hybrid dynamics in~\eqref{eq:hs_1obs}. 
\begin{lemma}
\label{lemma:finiteJumps}
For $\mathscr{H}$ in~\eqref{eq:hs_1obs}, each solution starting in $\mathcal{K}$ experiences no more than $3$ jumps.
\end{lemma}
Based on Lemmas~\ref{lemma:GAS_jumpless}-\ref{lemma:finiteJumps}, global asymptotic stability of $\A$ follows straightforwardly from~\cite[Thm.~31]{goebel2009hybrid} since the hybrid system in~\eqref{eq:hs_1obs} satisfies the Basic Assumptions \cite[p.~43]{goebel2009hybrid}, as proven in Lemma~\ref{lemma:hbc}, the set $\A$ is compact and has empty intersection with the jump set. 
%%%%%%%%%%%%%%%%%%%%%%%%%%%%%%%%%%
%%%%%%%%%%%%%%%%%%%%%%%%%%%%%%%%%%%%%%

Lastly, to prove item iii), let $\epsilon^\prime>\epsilon$. Select the parameter $\epsilon_h\in(\epsilon,\min(\epsilon^\prime,\sqrt{\epsilon\|c\|}))$ while all other hybrid controller parameters are selected as in Assumption~\ref{assumption:parameters}. Then this implies that the flow sets $\mathcal{F}_m, m\in\{-1,1\},$ of the avoidance mode are entirely contained in $\mathcal{B}_{\epsilon^\prime}(c)$. Therefore, as long as the state $x$ remains in $\mathbb{R}^n\setminus\mathcal{B}_{\epsilon^\prime}(c)$, solutions are enforced to flow only with the stabilizing mode $m=0$, which corresponds to the feedback law $u=-k_0x$.  
%%%%%%%%%%%%%%%%%%%%%%%%%%%%%%%%%%%%%%%%%%%%%%%%%%%%%%%%%%%%%%%%%%%%%%%%%%%%%%
%%%%%%%%%%%%%%%%%%%%%%%%%%%%%%%%%%%
\section{Numerical example}
\label{section:example}
\begin{figure}
    \centering
    \includegraphics[width=0.30\columnwidth]{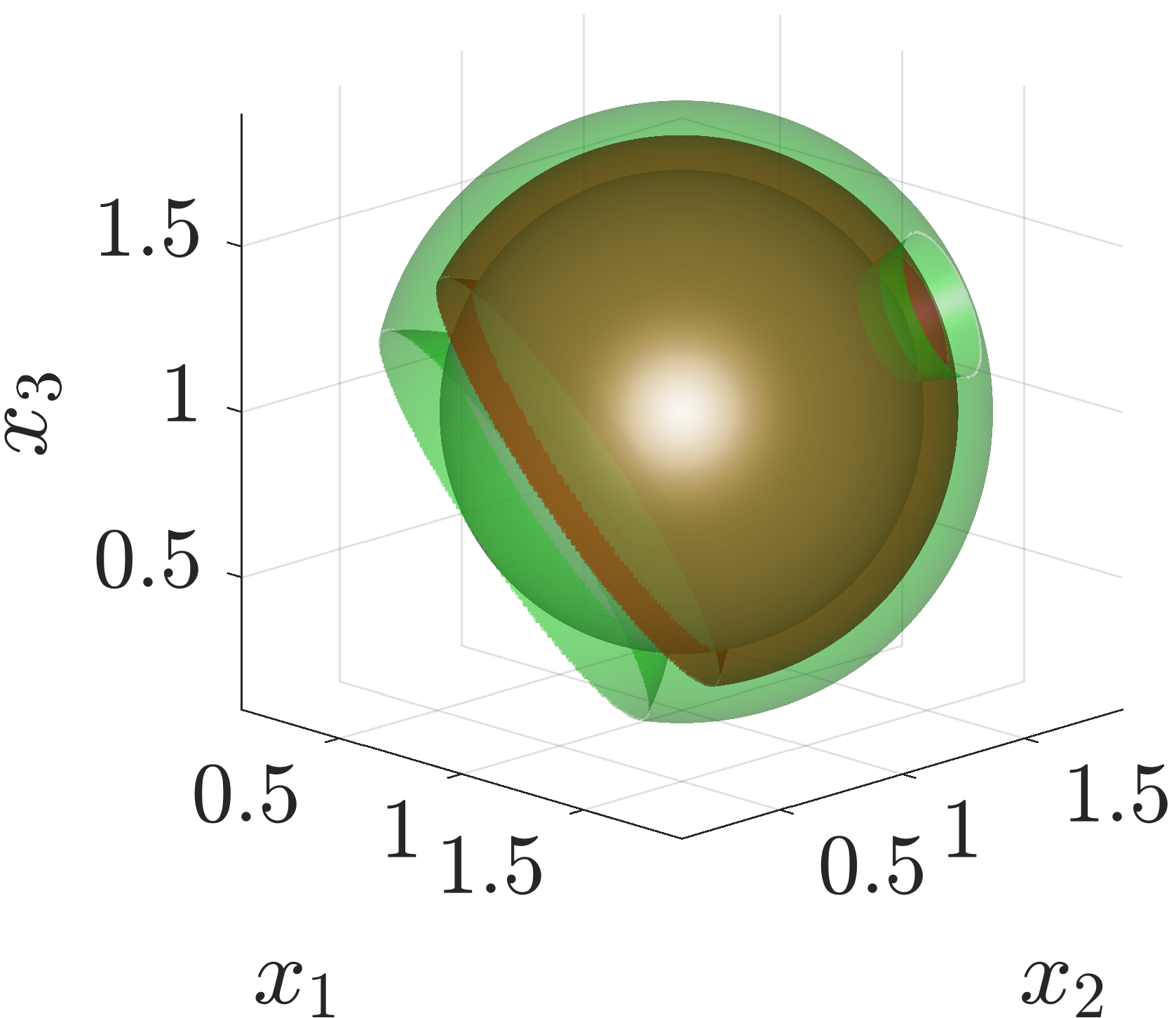}~~\includegraphics[width=0.25\columnwidth]{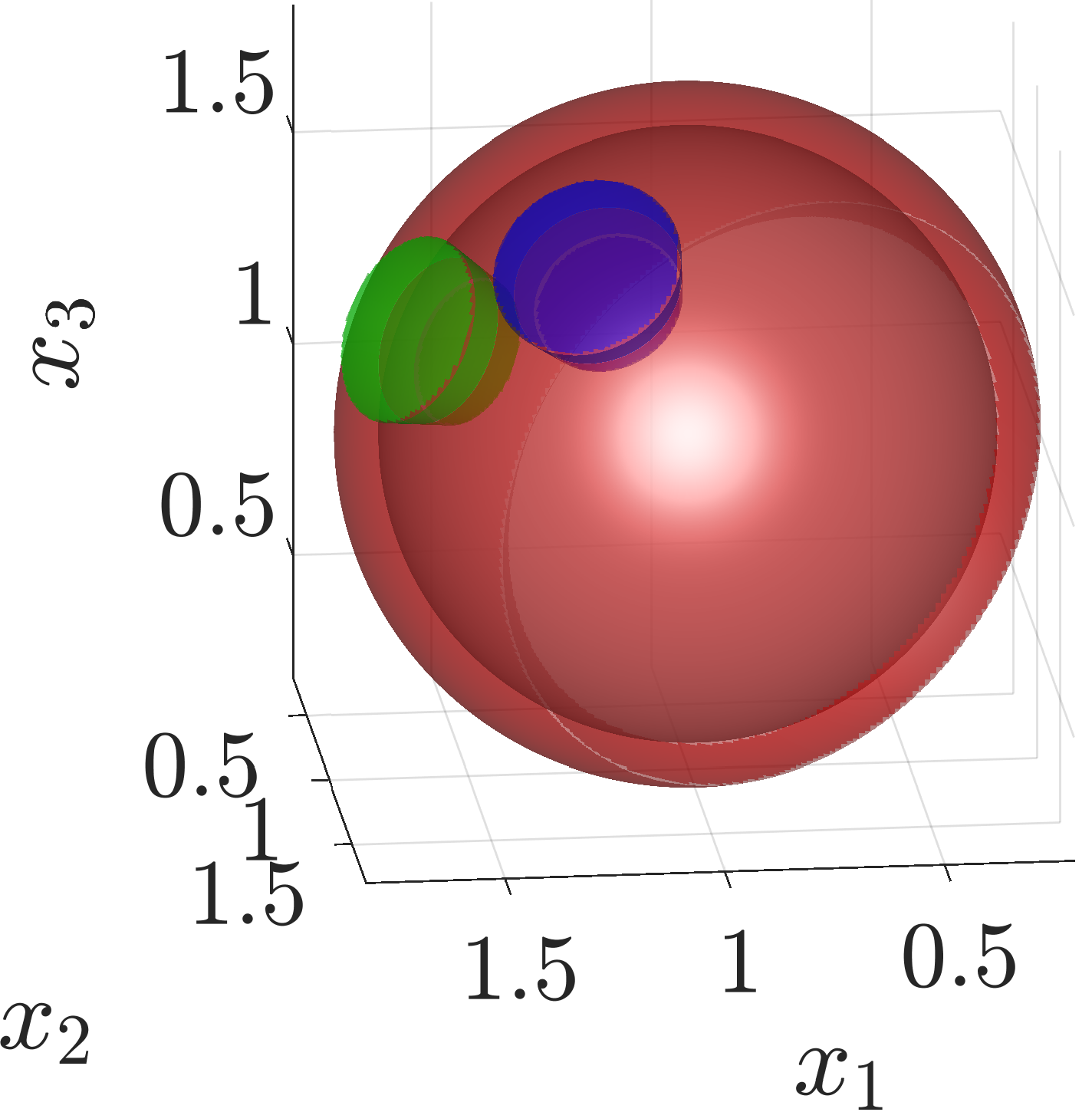}~~\includegraphics[width=0.30\columnwidth]{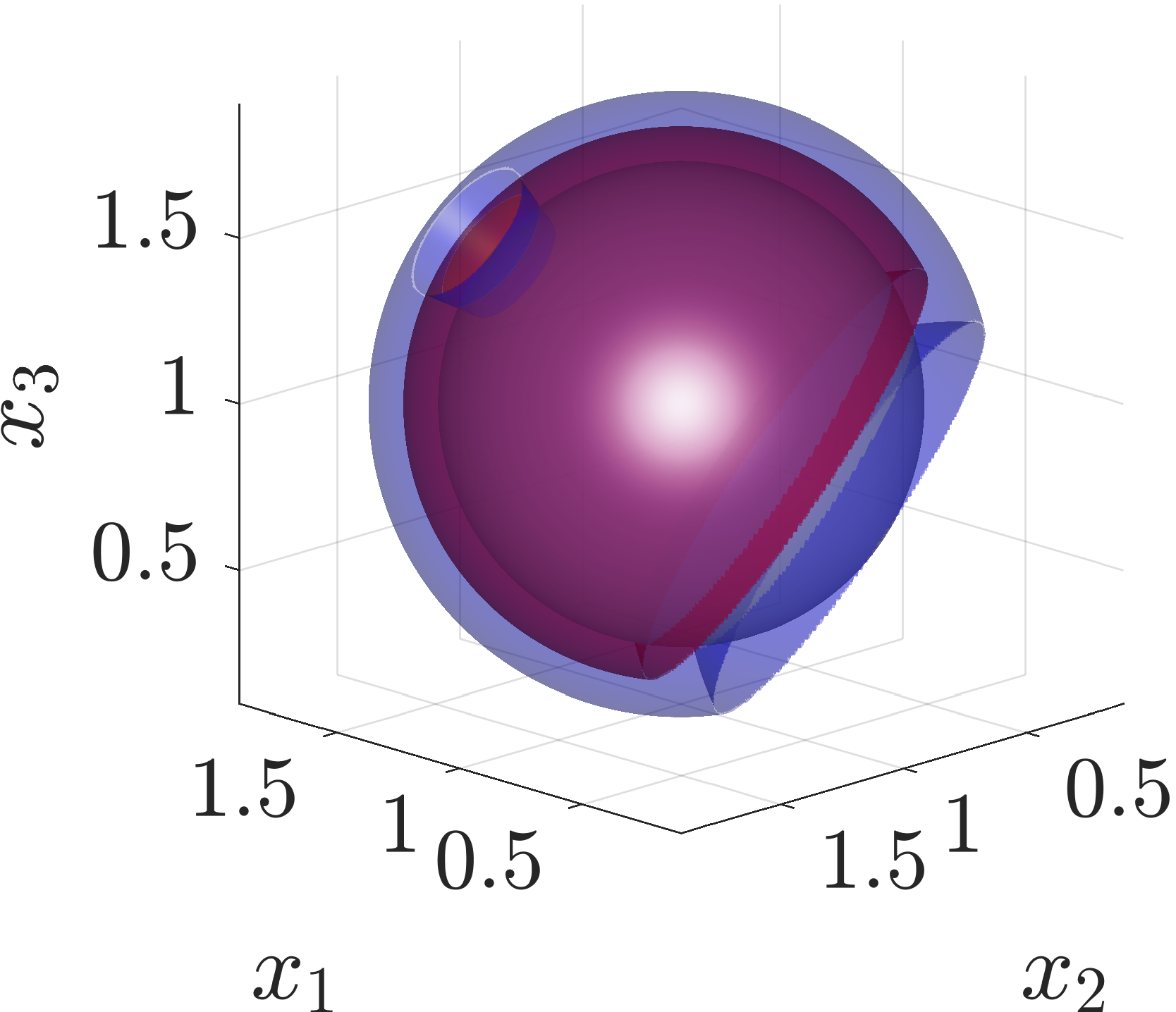}\\
    ~\\    
    \includegraphics[width=.97\columnwidth]{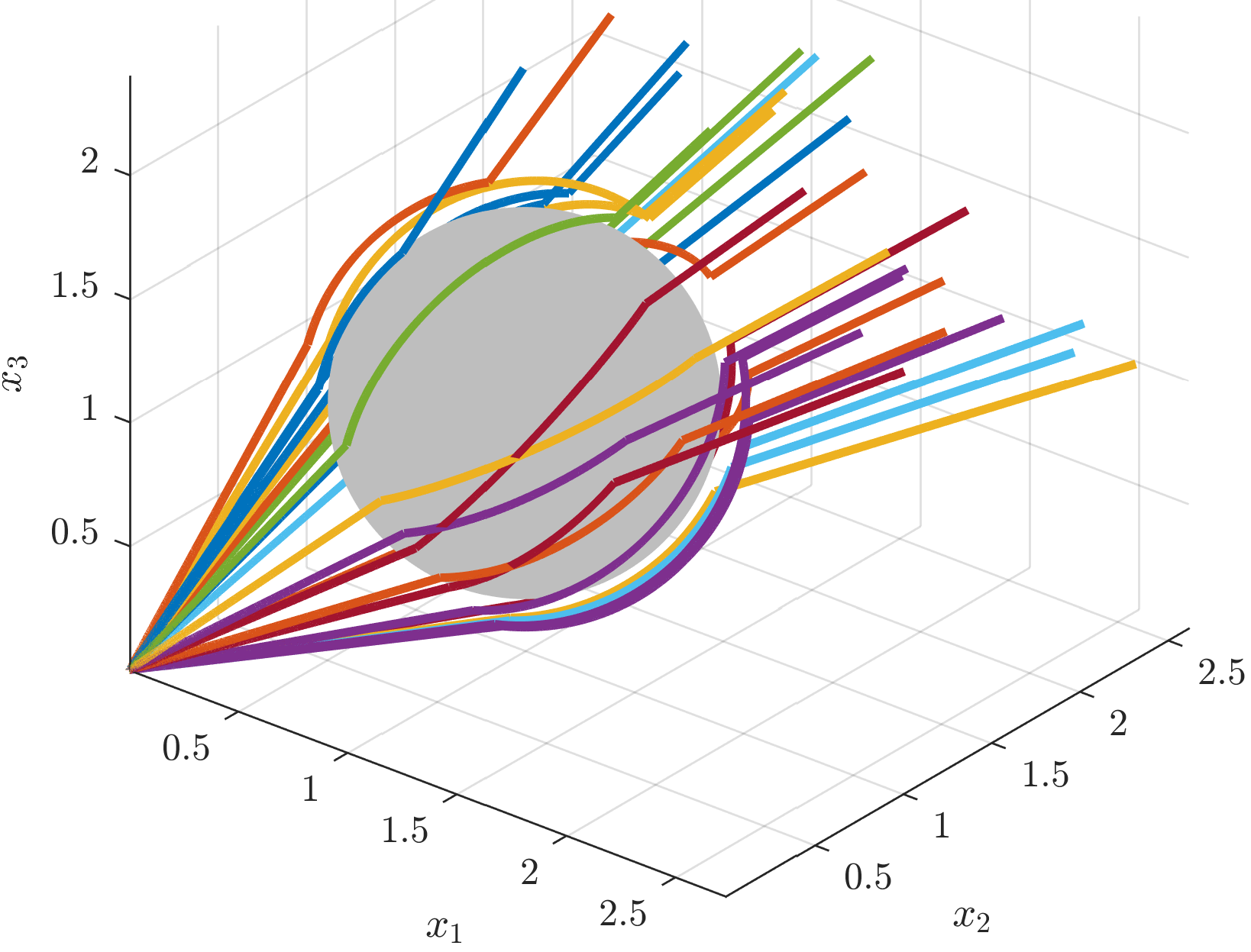}\\
    \includegraphics[width=.97\columnwidth]{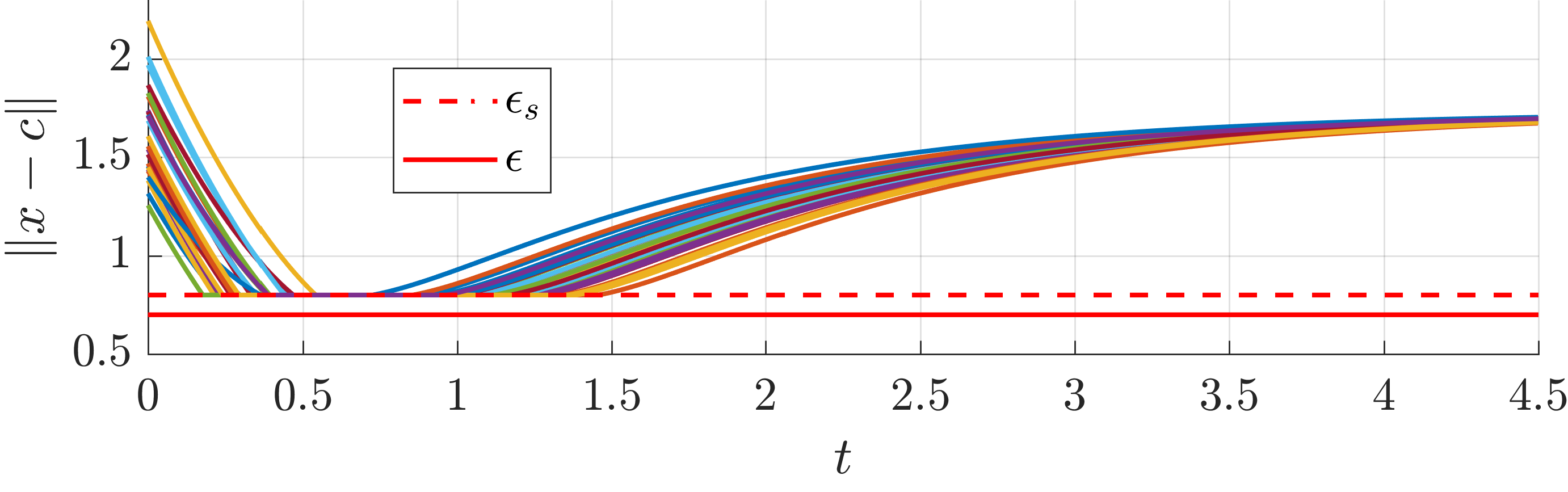}
    \caption{Top left: sets $\mathcal{F}_{-1}$ (green) and $\mathcal{J}_0$ (red) surrounding $\mathcal{B}_\epsilon(c)$ (grey). Top center: sets $\mathcal{J}_0$ (red), $\mathcal{J}_{-1} \cap \mathcal{H}(c,\epsilon,\epsilon_h, \mu)$ (green), and $\mathcal{J}_{1} \cap \mathcal{H}(c,\epsilon,\epsilon_h, \mu)$ (blue) surrounding $\mathcal{B}_\epsilon(c)$ (grey). Top right: sets $\mathcal{F}_1$ (blue) and $\mathcal{J}_0$ (red) surrounding $\mathcal{B}_\epsilon(c)$ (grey). 
    Middle: phase portrait of solutions with different initial conditions and $\mathcal{B}_\epsilon(c)$ (grey).
     Bottom: distance to the obstacle for the solutions and radii $\epsilon_s$, $\epsilon$ of $\mathcal{H}(c,\epsilon,\epsilon_s, 1/2)$, $\mathcal{B}_\epsilon(c)$.}
    \label{figure:construction}
\end{figure}

We illustrate our results through a three-dimensional example. The hybrid system in~\eqref{eq:hs_1obs} is fully specified by the following parameters. The obstacle has center $c=(1,1,1)$ and radius $\epsilon=0.700$. The controller gains are $k_m= 1$ for $m \in \{-1,0,1\}$. The parameters used in the construction of the flow and jump sets are $\epsilon_h = 0.901$, $\epsilon_s = 0.800$, $\mu=0.444$, $\theta= 0.276$, which satisfy Assumption~\ref{assumption:parameters}. To select a point $p_1\in\mathcal{C}^=_\leq(c,c,\theta)\setminus\{c\}$, we proceed as follows. Select $v\in\mathbb{S}^n$ such that $v^\top c=0$ and consider $\mathbf{R}(v,\theta)\in\mathbb{SO}(3)$, i.e., an orthogonal rotation matrix specified by axis $v$ and angle $\theta$. Then, we can verify that the point $p_1=(I_3-\mathbf{R}(v,\theta))c$ is a point on the cone $\mathcal{C}^=_\leq(c,c,\theta)$. By letting $v=(0,1,-1)$, we determine $p_1=(0.424,-0.155,-0.155)$ and $p_{-1}=(-0.348,0.231,0.231)$ as in~\eqref{eq:p-1}. We also select $\psi= 0.249$ and $\bar\psi = 0.266$, which satisfy Assumption~\ref{assumption:parameters}. 
Fig.~\ref{figure:construction} shows that the objectives posed in Section~\ref{section:problem} and proven in Theorem~\ref{theorem:invariance} are fulfilled. The top part of the figure illustrates the relevant sets. The middle part shows that the origin is globally asymptotically stable, and the control law matches the stabilizing one sufficiently away from the obstacle. The bottom part shows that the solutions are safe since they all stay away from the obstacle set $\mathcal{B}_\epsilon(c)$.

\appendix
\label{section:appendix}
All the lemmas are proven in this appendix.
% \section{Proof of Lemmas}\label{proof:lemmas}
%%%%%%%%%%%%%%%%%%%%%%%%%%%%%%%%%%%%%%%%%%%%%%%%%%
\subsubsection{Proof of Lemma \ref{lemma:cones}}
Let $x_i\in\mathcal{C}^{\leq}(c,v_i,\psi_i)\setminus\{c\}, i=1,2,$ and be otherwise arbitrary. Define then $z_i:=(x_i-c)/\|x_i-c\|\in\mathbb{S}^{n-1}$ for $i=1,2$. Hence, $z_i\in\mathcal{S}_i$ with $\mathcal{S}_i:=\mathcal{C}^{\leq}(0,v_i,\psi_i)\cap\mathbb{S}^{n-1}$, $i=1,2$. 
Since $z_i\in\mathcal{C}^{\leq}(0,v_i,\psi_i)$, either
$\mathbf{d}_{\mathbb{S}^{n-1}}(v_i, z_i) \le \psi_i$ (upper half cone) or $\mathbf{d}_{\mathbb{S}^{n-1}}(-v_i, z_i) \le \psi_i$ (lower half cone), for $i=1,2$. Consider all possible cases. 

If $\mathbf{d}_{\mathbb{S}^{n-1}}(v_i, z_i) \le \psi_i$ for both $i=1,2$, then it follows from the triangle inequality that
$
\theta = \mathbf{d}_{\mathbb{S}^{n-1}} (v_1,v_2) \leq \mathbf{d}_{\mathbb{S}^{n-1}}(v_1,z_1)+\mathbf{d}_{\mathbb{S}^{n-1}}(z_1,z_2)+\mathbf{d}_{\mathbb{S}^{n-1}}(v_2,z_2)\leq\mathbf{d}_{\mathbb{S}^{n-1}}(z_1,z_2)+\psi_1+\psi_2.
$
Hence, in view of the condition $\psi_1+\psi_2<\theta$,  $\mathbf{d}_{\mathbb{S}^{n-1}}(z_1,z_2)>0$. 

If, on the other hand, $\mathbf{d}_{\mathbb{S}^{n-1}}(-v_1, z_1) \le \psi_1$ and $\mathbf{d}_{\mathbb{S}^{n-1}}(v_2, z_2) \le \psi_2$, we have
$
\pi-\theta=\mathbf{d}_{\mathbb{S}^{n-1}}(-v_1,v_2)\leq\mathbf{d}_{\mathbb{S}^{n-1}}(-v_1,z_1)+\mathbf{d}_{\mathbb{S}^{n-1}}(z_1,z_2)+\mathbf{d}_{\mathbb{S}^{n-1}}(v_2,z_2) \leq\mathbf{d}_{\mathbb{S}^{n-1}}(z_1,z_2)+\psi_1+\psi_2.
$
Hence, in view of the condition $\theta<\pi-(\psi_1+\psi_2)$,  $\mathbf{d}_{\mathbb{S}^{n-1}}(z_1,z_2)>0$. 

The two cases of $\mathbf{d}_{\mathbb{S}^{n-1}}(-v_1, z_1) \le \psi_1$ and $\mathbf{d}_{\mathbb{S}^{n-1}}(-v_2, z_2) \le \psi_2$, and of $\mathbf{d}_{\mathbb{S}^{n-1}}(v_1, z_1) \le \psi_1$ and $\mathbf{d}_{\mathbb{S}^{n-1}}(-v_2, z_2) \le \psi_2$ lead analogously to the same conclusion. $\mathbf{d}_{\mathbb{S}^{n-1}}(z_1,z_2)>0$  implies that the sets $\mathcal{S}_1$ and $\mathcal{S}_2$ (and in turn $\mathcal{C}^{\leq}(c,v_i,\psi_i)\setminus\{c\}, i=1,2$) 
are disjoint.
%%%%%%%%%%%%%%%%%%%%%%%%%%%%%%%%%%%%%%%%%%%%%%%%%%%%%%%%%%%%%%%%%%%%%%%%%%%%%%%%%%%%%%%%%%%%%%%%
%
\subsubsection{Proof of Lemma \ref{lemma:p-1}}
By~\eqref{eq:p-1} and \eqref{eq:propLine3}, $p_{-1}-c=-\rho^\perp(c)(p_1-c)$. We can then show the claim. First, since $\rho^\perp(c)\pi^\theta(c)\rho^\perp(c)=\pi^\theta(c)$ (by \eqref{eq:def:piTheta}, \eqref{eq:propLine6}, \eqref{eq:propLine5}) and  $p_1\in\mathcal{C}^=(c,c,\theta)\setminus\{c\}$,
\begin{align*}
&(p_{-1}\!-\!c)^\top\!\pi^\theta(c)(p_{-1}\!-\!c)=(p_1\!-\!c)^\top\!\rho^\perp(c)\pi^\theta(c)\rho^\perp(c)(p_1\!-\!c)\\
&=(p_1-c)^\top\pi^\theta(c)(p_1-c)=0,
\end{align*}
i.e., $p_{-1}\in\mathcal{C}^=(c,c,\theta)\setminus\{c\}$. Second, by $p_1\in\mathcal{P}^\leq(c,c)$,
\begin{align*}
    c^\top(p_{-1}-c)=-c^\top\rho^\perp(c)(p_1-c)=c^\top(p_1-c)\leq 0,
\end{align*}
i.e., $p_{-1}\in\mathcal{P}^\leq(c,c)$. Therefore, $p_{-1}\in(\mathcal{C}^=(c,c,\theta)\setminus\{c\})\cap\mathcal{P}^\leq(c,c):=\mathcal{C}^=_\leq(c,c,\theta)\setminus\{c\}$.

\subsubsection{Proof of Lemma \ref{lemma:equilibria}}
Let $m$ be either $-1$ or $1$. The $\Longleftarrow$ implication is straightforward. As for the $\Longrightarrow$ implication, let $x\in\mathbb{R}^n\setminus\{c\}$ be such that $\pi^\perp(x-c)(x-p_m)=0$, which is equivalent to $\pi^\perp(x-c)(p_m-c)=0$. By the definition of the map $\pi^\perp(\cdot)$, one obtains $\|x-c\|^2(p_m-c)=(p_m-c)^\top(x-c)(x-c)$. However, $(p_m-c)^\top(x - c) \neq 0$, otherwise we would have $p_m = c$ (not true by~\eqref{eq:p-1} and Lemma~\ref{lemma:p-1}). Therefore, by letting $\lambda = \| x - c\|^2 /\big( (p_m-c)^\top (x-c) \big)$ in~\eqref{eq:def:line}, one deduces that $x\in\mathcal{L}(c,p_m-c)$.

\subsubsection{Proof of Lemma \ref{lemma:empty1}}
Let $m$ be either $-1$ or $1$. To deduce the claim, we prove first the relations:
\begin{subequations}
\label{eq:facts}
\begin{align}
\label{eq:fact1}
    &\mathcal{L}(c,p_m-c)\subset\mathcal{C}^=(c,c,\theta),\\
\label{eq:fact2}
    &\mathcal{L}(c,p_m-c)\setminus\{c\}\subset\mathcal{C}^<(c,p_m-c,\psi),\\
\label{eq:fact3}
    &\big(\mathcal{L}(c,p_m\!-\!c)\!\cap\!\mathcal{P}^\geq(c,p_m\!-\!c)\big)\subset\big(\mathcal{L}(c,p_m\!-\!c)\!\cap\!\mathcal{P}^\leq(c,c)\big),\\
\label{eq:fact4}
    &\mathcal{H}(c,\epsilon,\epsilon_h,\mu)\cap\mathcal{C}^=_\leq(c,c,\theta)=\emptyset.
\end{align}
\end{subequations}
As for~\eqref{eq:fact1}, let $x\in\mathcal{L}(c,p_m-c)$. Then there exists $\lambda$ such that $x-c=\lambda(p_m-c)$ and, hence,
\begin{align*}
(x-c)^\top\pi^{\theta}(c)(x-c)=\lambda^2(p_m-c)^\top\pi^{\theta}(c)(p_m-c)=0   
\end{align*}
since $p_m\in\mathcal{C}^=(c,c,\theta)$ by~\eqref{eq:p-1} and Lemma~\ref{lemma:p-1}, so \eqref{eq:fact1} is proven. As for~\eqref{eq:fact2}, let $x\in\mathcal{L}(c,p_m-c)\setminus\{c\}$. Then there exists $\lambda\neq 0$ such that $x-c=\lambda(p_m-c)$ and, hence,
\begin{align*}
&(x-c)\!^\top\!\pi^{\psi}(p_m\!-c)(x-c)\!=\!\lambda^2(p_m\!-c)\!^\top\!\pi^{\psi}\!(p_m\!-c)(p_m-c)\\
&=-\lambda^2\sin^2(\psi)\|p_m-c\|^2<0    
\end{align*}
by~\eqref{eq:def:piTheta}, \eqref{eq:propLine2}, \eqref{eq:propLine1}, so \eqref{eq:fact2} is proven. As for~\eqref{eq:fact3}, let $x\in\mathcal{L}(c,p_m-c)\cap\mathcal{P}^\geq(c,p_m-c)$. Then there exists $\lambda\geq 0$ such that $x-c=\lambda(p_m-c)$ and, hence,
\begin{align*}
    c^\top(x-c)=\lambda c^\top(p_m-c)=-\lambda\cos(\theta)\|c\|\|p_m-c\|\leq 0
\end{align*}
where we used $p_m\in\mathcal{C}^=_\leq(c,c,\theta)$ and $0<\theta<\theta_{\max}<\pi/2$ by Assumption~\ref{assumption:parameters}. Hence, one has $x\in\mathcal{P}^\leq(c,c)$, so \eqref{eq:fact3} is proven. As for~\eqref{eq:fact4}, let $x\in\mathcal{H}(c,\epsilon,\epsilon_h,\mu)$, then $x\in\mathcal{B}_{\epsilon_h}(c)$, $x\in\overline{\mathbb{R}^n\setminus\mathcal{B}_{\|\mu c\|}(\mu c)}$, and $x\in\overline{\mathbb{R}^n\setminus\mathcal{B}_{\epsilon}(c)}$ by~\eqref{eq:helmet}. So,
\begin{equation}
\label{eq:whenInH(epsilon_h,mu)}
\begin{aligned}
    &c^\top(c-x)=\frac{\|x-c\|^2+(1-\mu)^2\|c\|^2-\|x-\mu c\|^2}{2(1-\mu)}\\
    &\leq\frac{\epsilon_h^2+(1-\mu)^2\|c\|^2-\mu^2\|c\|^2}{2(1-\mu)}\!\!=\!\!\frac{\epsilon_h^2+\|c\|^2(1-2\mu)}{2(1-\mu)} \\
    &=\cos(\theta_{\max})\epsilon\|c\|\leq\cos(\theta_{\max})\|x-c\| \|c\|.
\end{aligned}    
\end{equation}
However, for all $z$, $z\in\mathcal{C}^=_\leq(c,c,\theta)$ is equivalent to $(z-c)^\top\pi^{\theta}(c)(z-c)=0$ and $c^\top(z-c)\leq 0$, i.e.,  $c^\top(z-c)=-\cos(\theta)\|z-c\|\|c\|<-\cos(\theta_{\max})\|z-c\| \|c\|$ by $\theta\in(0,\theta_{\max})$ in Assumption~\ref{assumption:parameters}. Then, by comparing with~\eqref{eq:whenInH(epsilon_h,mu)}, $x\notin\mathcal{C}^=_\leq(c,c,\theta)$, so \eqref{eq:fact4} is proven. Thanks to~\eqref{eq:facts}, the claim of the lemma is deduced as follows:
\begin{equation*}
    \begin{aligned}
        &\mathcal{F}_m\cap\mathcal{L}(c,p_m-c)\\
        &=(\mathcal{F}_m\cap\mathcal{L}(c,p_m-c)\cap\mathcal{P}^\geq(c,p_m-c))\\
        &\qquad\qquad \cup(\mathcal{F}_m\cap\mathcal{L}(c,p_m-c)\cap\mathcal{P}^<(c,p_m-c))\\
        &\overset{\eqref{eq:facts},\eqref{eq:def:half_cone}}{\subset}(\mathcal{F}_m\cap\mathcal{C}^=_\leq(c,c,\theta))\cup(\mathcal{F}_m\cap\mathcal{C}^<_<(c,p_m-c,\psi))\\
        &\overset{\eqref{eq:Fm}}{=}\mathcal{F}_m\cap\mathcal{C}^=_\leq(c,c,\theta)\\
        &\overset{\eqref{eq:Fm}}{=}\mathcal{H}(c,\epsilon,\epsilon_h,\mu)\cap\mathcal{C}_\leq^\geq(c,p_m-c,\psi)\cap\mathcal{C}^=_\leq(c,c,\theta)=\emptyset.
    \end{aligned}
\end{equation*}
%%%%%%%%%%%%%%%%%%%%%%%%%%%%%%%%%%%%%%%%%%%%%%%%%%%%%%%%%%%%%%%%%%%%%%%%%%%%%%%%%%%%%%%%%%%%%%%%%%%%%%%%%%%%%%%%%%%%%%%%%%%%%%%%%%%%%%%%%%%%%%%%%%%%%%%%%%%%%%%%%%%%
%%%%%%%%%%%%%%%%%%%%%%%%%%%%%%
\subsubsection{Proof of Lemma \ref{lemma:hbc}}
$\mathcal{F}$ and $\mathcal{J}$ are closed subsets of $\real^{n}\times\{-1,0,1\}$. $\mathbf{F}$ is a continuous function in $\mathcal F$ (hence, it is outer semicontinuous and locally bounded relative to $\mathcal{F}$, $\mathcal{F} \subset \dom \mathbf{F}$, and $\mathbf{F}(x,m)$ is convex for every $(x,m)\in\mathcal{F}$). $\mathbf{J}$ has a closed graph in $\mathcal{J}$, is locally bounded relative to $\mathcal{J}$ and is nonempty on $\mathcal{J}$. In particular, let us show that $\mathbf{M}(x,0) \neq \emptyset$ for all $x \in \mathcal{J}_0$. 

We preliminarily show that $\cap_{m=-1,1}\mathcal{C}^{\leq}(c,p_m-c,\bar\psi)=\{ c \}$. Let $v_m=(p_m-c)/\|p_m-c\|$, and substitute in 
\begin{align*}
v_1^\top v_{-1}&=\frac{(p_1-c)^\top(p_{-1}-c)}{\|p_1-c\|\|p_{-1}-c\|}=\frac{-(p_1-c)^\top\rho^\perp(c)(p_1-c)}{\|p_1-c\|\|\rho^\perp(c)(p_1-c)\|}\\
                &=-\frac{(p_1-c)^\top(2\pi^\theta(c)-\cos(2\theta)I_n)(p_1-c)}{\|p_1-c\|\|p_1-c\|}\\ 
                &=\cos(2\theta)\frac{(p_1-c)^\top(p_1-c)}{\|p_1-c\|\|p_1-c\|}=\cos(2\theta)
\end{align*}
where we have used, in this order, the facts that $\rho^\perp(c)=2\pi^\theta(c)-\cos(2\theta)I_n$, $\rho^\perp(c)\rho^\perp(c)=I_n$ and $(p_1-c)^\top\pi^\theta(c)(p_1-c)=0$ (since $p_1\in\mathcal{C}^=(c,c,\theta)$ is implied by~\eqref{eq:p-1}).
Then, by Lemma~\ref{lemma:cones} and $2\bar\psi<2\theta<\pi-2\bar\psi$ (from~\eqref{ineq:parameters} and \eqref{eq:psi_max}, $\bar \psi < \min(\theta,\pi/2-\theta)$), $\cap_{m=-1,1}\mathcal{C}^{\leq}(c,p_m-c,\bar\psi)=\{ c \}$. Hence, it can be shown by a contradiction argument that $\cup_{m=-1,1}\mathcal{C}^{\geq}(c,p_m-c,\bar\psi)=\mathbb{R}^n$. Therefore, in view of \eqref{eq:hs_1obs:M(x,0)}, the set $\mathbf{M}(x,0)$ is nonempty.

Finally, $\mathbf{M}(x,0)$ has a closed graph since the construction in~\eqref{eq:hs_1obs:M(x,0)} allows $\mathbf{M}$ to be set-valued whenever $x\in\cap_{m=-1,1}\mathcal{C}^{\geq}(c,p_m-c,\bar\psi)\cap \mathcal{J}_0$.

%%%%%%%%%%%%%%%%%%%%%%%%%%%%%
\subsubsection{Proof of Lemma \ref{lemma:GAS_jumpless}}
Consider the Lyapunov function
\begin{equation}
\label{eq:V}
    \mathbf{V}(x,m):= m^2/2 + \| x - p_m \|^2/2,
\end{equation}
with $p_0:=0$ and $p_{m}$ ($m\in\{-1,1\}$) defined in \eqref{eq:p-1}. One has $\mathbf{V}(x,m)=0$ for all $(x,m) \in \A$ in~\eqref{eq:KandA}, $\mathbf{V}(x,m)>0$ for all $(x,m) \notin \A$, and is radially unbounded relative to $\mathcal{F} \cup \mathcal{J}$. Straightforward computations show that
\begin{align*}
&     \langle \nabla \mathbf{V} (x,0), \mathbf{F}(x,0) \rangle   =-k_0 x^\top x < 0 \quad \forall x \in \mathcal{F}_0 \setminus \{0 \}   
\\
& \begin{aligned}
    &\langle \nabla \mathbf{V} (x,m),  \mathbf{F}(x,m) \rangle = -k_m (x-p_m)^\top \pi^\perp(x-c) (x-p_m) \\
    &  =\! -k_m \| \pi^\perp\! (x-c) (x-p_m) \|^2 \! < \! 0\quad \forall m \in\{-1,1\}, x \in \mathcal{F}_m.
\end{aligned} %\label{eq:Vdot,mNotZero}
\end{align*}
The last inequality follows from projection matrices being positive semidefinite and Lemma~\ref{lemma:equilibria}, which implies that it cannot be $\langle \nabla \mathbf{V} (x,m),  \mathbf{F}(x,m) \rangle = 0$ for $m \in\{-1,1\}$ and all $x \in \mathcal{F}_m$ since $\mathcal{L}(c,p_m - c)$ is excluded from $\mathcal{F}_m$ by Lemma~\ref{lemma:empty1}. All the above conditions satisfied by $\mathbf{V}$ suffice to conclude global asymptotic stability of $\A$ for $\mathscr{H}^0$ since $\A$ is compact and $\mathscr{H}^0$ satisfies \cite[Ass.~6.5]{goebel2012hybrid}.
%%%%%%%%%%%%%%%%%%%%%%%%%%%%%%%%%%%%%%%%%
\subsubsection{Proof of Lemma \ref{lemma:finiteJumps}}
We prove, case by case, that the number of jumps, denoted $N$, does not exceed $3$.

%{\customlabel{lemma:finiteJumps:m=0}{\textit{(i)}} \ref{lemma:finiteJumps:m=0}
    \textit{(i)} {\bf Case $m(0,0)=0$.}
    Let us define the disjoint sets
\begin{align}
    &\mathcal{R}_a:=\mathcal{C}_\geq^{\leq}(0,c,\gamma)\setminus\mathcal{B}_{\|c/2\|}(c/2)\setminus\mathcal{B}_{\epsilon_s}(c), \label{eq:Ra}\\
    &\mathcal{R}_b:=\mathcal{F}_0 \setminus(\mathcal{R}_a\cup\mathcal{J}_0)
\end{align}
with $\cos(\gamma):=\sqrt{1-{\epsilon_s^2}/{\|c\|^2}}$ (well-defined by Assumption~\ref{assumption:obstacle} and \eqref{ineq:eps_h,eps_s}). Note that $\mathcal{R}_a\cup\mathcal{R}_b\cup\mathcal{J}_0=\overline{\mathbb{R}^n\setminus\mathcal{B}_\epsilon(c)}$.
%}

%{\customlabel{lemma:finiteJumps:m=0,Rb}{\textit{(i.1)}} \ref{lemma:finiteJumps:m=0,Rb}
\textit{(i.1)}
$x(0,0)\in\mathcal{R}_b$: Solutions can only flow. Consider then the jumpless hybrid system in $\real^n$ with data $(-k_0x,\mathcal{R}_b,\emptyset,\emptyset)$ and let us show that maximal solutions are complete. Since finite escape times are excluded, it is sufficient (by, e.g., \cite[Prop.~2.10]{goebel2012hybrid}) to show that the viability condition $\{-k_0 x\}\subset\mathbf{T}_{\overline{\mathcal{R}_b}}(x)$ holds for all $x\in\partial\mathcal{R}_b$, with 
\begin{align*}
    \partial\mathcal{R}_b&=\big( \partial\mathcal{B}_\epsilon(c)\cap\mathcal{B}_{\|c/2\|}(c/2)\big)
    \cup
    \big(\partial\mathcal{B}_{\|c/2\|}(c/2)\cap\mathcal{B}_{\epsilon_s}(c)\\
    &\setminus\mathcal{B}_\epsilon(c)\big) 
    \cup
    \big(\mathcal{C}^=_\geq(0,c,\gamma)\setminus\mathcal{B}_{\|c/2\|}(c/2)\big)
\end{align*}
and $\mathbf{T}_{\overline{\mathcal{R}_b}}(x)$ in the following table.
\begin{table}[h!]
\begin{tabularx}{\columnwidth}{Xl}
\toprule
Set to which $x$ belongs & $\mathbf{T}_{\overline{\mathcal{R}_b}}(x)$\\
\midrule
$\partial\mathcal{B}_\epsilon(c)\cap\mathcal{B}^\circ_{\|c/2\|}(c/2)$ & $\mathcal{P}^\geq(0,x-c)$\\
$(\partial\mathcal{B}_{\|c/2\|}\!(c/2) \cap \mathcal{B}^\circ_{\epsilon_s}\!(c))\!\setminus\!\mathcal{B}_\epsilon(c)$ & $\mathcal{P}^\leq(0,x-c/2)$\\
$\mathcal{C}^=_\geq(0,c,\gamma)\setminus\mathcal{B}_{\|c/2\|}(c/2)$ & $\mathcal{P}^\geq(0,\pi^\gamma(c)x)$\\
$\partial\mathcal{B}_{\epsilon}(c)\cap\partial\mathcal{B}_{\|c/2\|}(c/2)$ & $\mathcal{P}^\geq(0,x-c)\cap\mathcal{P}^\leq(0,x-c/2)$ \\
$\partial\mathcal{B}_{\|c/2\|}(c/2)\cap\mathcal{C}^=_\geq(0,c,\gamma)$ & $\mathcal{P}^\geq(0,\pi^\gamma(c)x)\cup\mathcal{P}^\leq(0,x-c/2)$\\
\bottomrule
\end{tabularx}
\end{table}

Let $z:=-k_0x$ and let us show that $z\in\mathbf{T}_{\overline{\mathcal{R}_b}}(x)$ for all $x \in \partial \mathcal{R}_b$. If $x\in\mathcal{B}_{\|c/2\|}(c/2)$, then $z^\top(x-c)=-k_0 x^\top(x-c)\geq 0$, hence $z\in\mathcal{P}^\geq(0,x-c)$. If $x\in\partial\mathcal{B}_{\|c/2\|}(c/2)$, then $z^\top (x-c/2)=-k_0x^\top(x-c/2)=-k_0x^\top c/2=-k_0\|x\|^2/2 \le 0$, hence $z\in\mathcal{P}^\leq(0,x-c/2)$. Finally, if $x\in\mathcal{C}^=_\geq(0,c,\gamma)$, then $z^\top\pi^\gamma(c)x=-k_0 x^\top\pi^\gamma(c)x=0$ implying that $z\in\mathcal{P}^\ge(0,\pi^\gamma(c)x)$, where $\pi ^\gamma(c) x$ is a normal vector to $\mathcal{C}^=_\geq(0,c,\gamma)$ at $x$. By combining these cases and inspecting the previous table, the above viability condition holds for all $x \in \partial \mathcal{R}_b$, hence solutions are complete. Therefore, $N=0$ for each solution with this initial condition.
%}

%{\customlabel{lemma:finiteJumps:m=0,Ra}{\textit{(i.2)}} \ref{lemma:finiteJumps:m=0,Ra}
\textit{(i.2)} $x(0,0)\in\mathcal{R}_a$: We argue that $\mathcal{J}_0$ is reached in finite time. Let us preliminarily show that
\begin{equation}
\label{eq:RaBndry}
\partial\mathcal{B}_{\|c/2\|}(c/2)\cap\mathcal{C}_\geq^{<}(0,c,\gamma)\subset \mathcal{B}^\circ_{\epsilon_s}(c).    
\end{equation}
Let $x\in\partial\mathcal{B}_{\|c/2\|}(c/2)\cap\mathcal{C}_{\geq}^<(0,c,\gamma)$. Since $x\in\partial\mathcal{B}_{\|c/2\|}(c/2)$, one has $\|x-c/2\|^2=\|c/2\|^2$, i.e., $\|x\|^2=c^\top x$. Besides, since $x\in\mathcal{C}_{\geq}^<(0,c,\gamma)$, one has $c^\top x=\|x\|^2>\cos(\gamma)\|x\|\|c\|$, i.e.,  $-\|x\|^2<\epsilon_s^2 -\|c\|^2$ by the definition of $\cos(\gamma)$ in~\textit{(i)}.
%\ref{lemma:finiteJumps:m=0}
By $\| x \|^2 = c^\top x$ and the last bound, we have $\|x-c\|^2=\|c\|^2-\|x\|^2<\epsilon_s^2$, i.e., $x\in\mathcal{B}^\circ_{\epsilon_s}(c)$. Therefore, by~\eqref{eq:RaBndry} and \eqref{eq:Ra}, it can only be $\partial\mathcal{R}_a  = \big( \mathcal{C}^=_\geq (0,c,\gamma )  \setminus  \mathcal{B}_{\| c/2 \|} (c/2)\big) \cup\big(\partial \mathcal{B}_{\epsilon_s} (c)  \setminus  \mathcal{B}^\circ_{\| c/2 \|} (c/2)\big)$. 

Then, note that maximal solutions to~\eqref{eq:hs_1obs} with the current initial condition are complete by item i), previously proven. Since $\mathbf{V}(x,0):=\| x \|^2/2$ in~\eqref{eq:V} is strictly decreasing along the flow in $\overline{\mathcal{R}_a}$ and is bounded from below, such complete solutions cannot flow indefinitely in $\mathcal{R}_a \times \{ 0 \}$ and must leave this set in finite time. On the other hand, they cannot leave through $\mathcal{C}^=_\geq(0,c,\gamma)\setminus \mathcal{B}_{\| c/2 \|}(c/2)$. Indeed, for all $x\in\mathcal{C}^=_\geq(0,c,\gamma)$, $(-k_0 x)^\top\pi^\gamma(c)x=0$ and thus $\{-k_0x\}\in\mathcal{P}^=(0,\pi^\gamma(c)x)\subset\mathcal{P}^\leq(0,\pi^\gamma(c)x)$ which is the tangent cone of $\mathcal{R}_a$ at $x$ ($\pi^\gamma(c)x$ is defined in item~\textit{(i.1)}).
%\ref{lemma:finiteJumps:m=0,Rb}). 
It follows that solutions must leave $\mathcal{R}_a$ through $\partial\mathcal{B}_{\epsilon_s}(c)\setminus\mathcal{B}^\circ_{\|c/2\|}(c/2)\subset\mathcal{J}_0$, that is, they reach $\mathcal{J}_0$ in finite time. From there, the analysis boils down to that in item~\textit{(i.3)}.
%\ref{lemma:finiteJumps:m=0,J0}. 
Therefore, $N = 2$ for each solution with this initial condition.
%}

%{\customlabel{lemma:finiteJumps:m=0,J0}{\textit{(i.3)}} \ref{lemma:finiteJumps:m=0,J0}
\textit{(i.3)} $x(0,0)\in\mathcal{J}_0$: According to the jump map,  $m(0,1)= m^\prime$ for some $ m^\prime\in \{ -1, 1\}$ and the jump map in~\eqref{eq:hs_1obs:M(x,0)} ensures  $x(0,0)=x(0,1) \in \mathcal{C}^\ge (c,p_{m^\prime}-c, \bar \psi)$. Therefore, since we selected $\bar \psi>\psi$ in~\eqref{ineq:parameters}, one has $x(0,1) \in \mathcal{C}^\ge (c,p_{m^\prime}-c, \bar \psi)\cap \mathcal{J}_0 \subset \mathcal{C}^>(c,p_{m^\prime}-c,\psi)\cap\mathcal{J}_0\subset\mathcal{F}_{ m^\prime} \setminus \mathcal{J}_{ m^\prime}$. Hence, $x(0,1) \in \mathcal{F}_{ m^\prime} \setminus \mathcal{J}_{ m^\prime}$, thereby excluding a further consecutive jump. We show in item~\textit{(ii.2)}
%\ref{lemma:finiteJumps:mBar,FmBar}
that after a flow, one jump is experienced. Therefore, $N = 2$ for each solution with this initial condition.
%}

%{\customlabel{lemma:finiteJumps:mBar}{\textit{(ii)}} \ref{lemma:finiteJumps:mBar}
\textit{(ii)} {\bf Case $m(0,0)=\bar m\in\{-1,1\}$.}
%}

%{\customlabel{lemma:finiteJumps:mBar,JmBar}{\textit{(ii.1)}} \ref{lemma:finiteJumps:mBar,JmBar}
\textit{(ii.1)} $x(0,0)\in\mathcal{J}_{\bar m}$: According to the jump map, one has $m(0,1)=0$ and the cases \textit{(i.1)}, \textit{(i.2)}, or \textit{(i.3)}
%\ref{lemma:finiteJumps:m=0,Rb}, \ref{lemma:finiteJumps:m=0,Ra} or \ref{lemma:finiteJumps:m=0,J0} 
can occur. Therefore $N\le 3$ for each solution with this initial condition.
%}

%{\customlabel{lemma:finiteJumps:mBar,FmBar}{\textit{(ii.2)}} \ref{lemma:finiteJumps:mBar,FmBar}
\textit{(ii.2)} $x(0,0)\in \mathcal{F}_{\bar m}\setminus\mathcal{J}_{\bar m}$. 
An argument similar to that in~\textit{(i.2)}
%\ref{lemma:finiteJumps:m=0,Ra} 
concludes that solutions to~\eqref{eq:hs_1obs} with this initial condition must leave $\mathcal{F}_{\bar m}\setminus\mathcal{J}_{\bar m}$ in finite time.
Indeed, solutions are complete by Theorem~\ref{theorem:invariance} and $\mathbf{V}(x,\bar m)$ in~\eqref{eq:V} is strictly decreasing along the flow in $\mathcal{F}_{\bar m}$ by the proof in Lemma~\ref{lemma:GAS_jumpless} and bounded from below, so solutions cannot flow indefinitely in $\mathcal{F}_{\bar m}$.
Then, by similar arguments as in the previously proven item i) of Theorem~\ref{theorem:invariance}, solutions can reach in finite time only the set $\mathcal{L}_{\bar m}$ (defined there, above~\eqref{eq:tgConeBndFm}). However, $\mathcal{L}_{\bar m} \subset \mathcal{R}_b$, and we have shown in item~\textit{(i.1)}
%\ref{lemma:finiteJumps:m=0,Rb} 
that no jumps are experienced in $\mathcal{R}_b$. Therefore, $N=1$ for each solution with this initial condition.
%}

Because all the possible cases for $x$ and $m$ are covered without circularity, we conclude then that each solution starting in $\mathcal{K}$ experiences no more than 3 jumps.

\bibliographystyle{IEEEtran}
\bibliography{IEEEabrv,Bibliography_multiagent}
\end{document}